\documentclass[aps,prd,twocolumn,showkeys,amsmath,amssymb]{revtex4}
\usepackage{graphicx}
\usepackage{epstopdf}
\usepackage{multirow}
\usepackage[colorlinks,citecolor=blue,anchorcolor=red,menucolor=red,linkcolor=red,filecolor=red,runcolor=red,urlcolor=blue,frenchlinks=red]{hyperref}
\usepackage{soul}
\usepackage{extarrows}

\allowdisplaybreaks[3]

\begin{document}
%=====================================================================================
%=====================================================================================
\title{Decay behaviors of the fully-bottom and fully-charm tetraquark states}
%=====================================================================================
%=====================================================================================
%

\author{Hua-Xing Chen}
\email{hxchen@seu.edu.cn}
\affiliation{School of Physics, Southeast University, Nanjing 210094, China}
\author{Yi-Xin Yan}
\email{yxyan@seu.edu.cn}
\affiliation{School of Physics, Southeast University, Nanjing 210094, China}
\author{Wei Chen}
\email{chenwei29@mail.sysu.edu.cn}
\affiliation{School of Physics, Sun Yat-Sen University, Guangzhou 510275, China}

\begin{abstract}
We study the decay behaviors of the fully-bottom tetraquark states within the diquark-antidiquark picture, and calculate their relative branching ratios through the Fierz rearrangement. Our results suggest that the $C=+$ states can be searched for in the $\mu^+ \mu^- \Upsilon(1S)$ and $\mu^+ \mu^- \Upsilon(2S)$ channels with the relative branching ratio $\mathcal{B}(X \to \mu\mu \Upsilon(2S)) / \mathcal{B}(X \to \mu \mu \Upsilon(1S)) \approx 0.4$. Our results also suggest that the $C=-$ states can be searched for in the $\mu^+ \mu^- \eta_b(1S)$ and $\mu^+ \mu^- \eta_b(2S)$ channels with the similar relative branching ratio $\mathcal{B}(X \to \mu\mu \eta_b(2S)) / \mathcal{B}(X \to \mu \mu \eta_b(1S)) \approx 0.4$. We also reanalysis the fully-charm tetraquark states, and study the $X(6900)$ decay into the $J/\psi \psi(2S)$ channel to obtain the relative branching ratio $\mathcal{B}(X \to J/\psi \psi(2S)) / \mathcal{B}(X \to J/\psi J/\psi) \approx 0.1$.
\end{abstract}
\pacs{12.39.Mk, 12.38.Lg, 12.40.Yx}
\keywords{exotic hadron, fully-bottom tetraquark, fully-charm tetraquark, Fierz rearrangement}
\maketitle
\pagenumbering{arabic}

\section{Introduction}
\label{sec:intro}

In 2020 the LHCb collaboration reported their observation of two exotic structures in the di-$J/\psi$ invariant mass spectrum~\cite{LHCb:2020bwg}, {\it i.e.}, a broad structure ranging from 6.2 to 6.8~GeV and a narrow structure at around 6.9~GeV. They described the latter as a resonance with the Breit-Wigner lineshape, whose mass and width were measured to be
\begin{eqnarray}
X(6900) &:& M = 6905 \pm 11 \pm 7 {\rm~MeV} \, ,
\\ \nonumber && \Gamma = 80 \pm 19 \pm 33 {\rm~MeV} \, .
\end{eqnarray}
These values were obtained under the assumption that no interference with the non-resonant single-parton scattering continuum is present. Assuming that the continuum interferes with the broad structure, the above values were shifted to be
\begin{eqnarray}
X(6900) &:& M = 6886 \pm 11 \pm 11 {\rm~MeV} \, ,
\\ \nonumber && \Gamma = 168 \pm 33 \pm 69 {\rm~MeV} \, .
\end{eqnarray}
The above two structures are good candidates for the fully-charm tetraquark states, and their observation immediately attracted much attention from the particle physics community~\cite{liu:2020eha,Tiwari:2021tmz,Lu:2020cns,Faustov:2020qfm,Zhang:2020xtb,Li:2021ygk,Bedolla:2019zwg,Weng:2020jao,Liu:2021rtn,Giron:2020wpx,Karliner:2020dta,Zhao:2020zjh,Mutuk:2021hmi,Wang:2021kfv,Wang:2020ols,Ke:2021iyh,Zhu:2020xni,Jin:2020jfc,Yang:2021hrb,Albuquerque:2020hio,Albuquerque:2021erv,Wu:2022qwd,Asadi:2021ids,Yang:2020wkh,Huang:2020dci,Feng:2020riv,Huang:2021vtb,Ma:2020kwb,Maciula:2020wri,Goncalves:2021ytq,Wang:2020gmd,Esposito:2021ptx,Zhuang:2021pci,Zhao:2020nwy,Becchi:2020uvq,Andrade:2022rbn,Lansberg:2020ejc,Sonnenschein:2020nwn,Zhu:2020snb,Wan:2020fsk,Gordillo:2020sgc,Liu:2020tqy,Majarshin:2021hex,Sombillo:2021rxv,Kuang:2022vdy,Yang:2020rih,Yang:2021zrc,Wang:2021taf,Wang:2021mma}. We refer to our recent review~\cite{Chen:2022asf} as well as the reviews~\cite{Chen:2016qju,Liu:2019zoy,Lebed:2016hpi,Esposito:2016noz,Hosaka:2016pey,Guo:2017jvc,Ali:2017jda,Olsen:2017bmm,Karliner:2017qhf,Bass:2018xmz,Brambilla:2019esw,Guo:2019twa,Ketzer:2019wmd,Yang:2020atz,Roberts:2021nhw,Fang:2021wes,Jin:2021vct,JPAC:2021rxu,Meng:2022ozq,Mai:2022eur,Maiani:2022psl} and the reports~\cite{Maiani:2020pur,Chao:2020dml,Richard:2020hdw} for their detailed discussions.
Especially, some theorists reanalysed the LHCb data on the di-$J/\psi$ spectrum~\cite{LHCb:2020bwg} and proposed the existence of more structures, {\it e.g.}, the authors of Ref.~\cite{Wang:2020wrp} reproduced three peak structures at near 6.5, 6.9, and 7.3~GeV, while the authors of Ref.~\cite{Dong:2020nwy} proposed the existence of a near-threshold state in the di-$J/\psi$ system at near 6.2~GeV. We refer to Refs.~\cite{Wang:2020tpt,Dong:2020hxe,Liang:2021fzr,Nefediev:2021pww,Gong:2020bmg,Dong:2021lkh,Guo:2020pvt,Cao:2020gul,Wang:2022jmb,Zhou:2022xpd} and the reviews~\cite{Chen:2022asf,Mai:2022eur} for more discussions.

Very recently, the CMS and ATLAS collaborations also investigated the di-$J/\psi$ invariant mass spectrum, and both of them confirmed the existence of the $X(6900)$~\cite{CMS,ATLAS}. Besides, the CMS collaboration observed two new structures, the $X(6600)$ and $X(7200)$, in the di-$J/\psi$ invariant mass spectrum. Their masses and widths were measured to be~\cite{CMS}:
\begin{eqnarray}
X(6600) &:& M = 6552 \pm 10 \pm 12 {\rm~MeV} \, ,
\\ \nonumber && \Gamma = 124 \pm 29 \pm 34 {\rm~MeV} \, ;
\\ X(6900) &:& M = 6927 \pm 9 \pm 5 {\rm~MeV} \, ,
\\ \nonumber && \Gamma = 122 \pm 22 \pm 19 {\rm~MeV} \, ;
\\ X(7200) &:& M = 7287 \pm 19 \pm 5 {\rm~MeV} \, ,
\\ \nonumber && \Gamma = 95 \pm 46 \pm 20 {\rm~MeV} \, .
\end{eqnarray}
The ATLAS collaboration investigated the di-$J/\psi$ invariant mass spectrum, and their best fit was performed with three interfering resonances, whose masses and widths were measured to be~\cite{ATLAS}:
\begin{eqnarray}
X(6200) &:& M = 6.22 \pm 0.05 ^{+0.04}_{-0.05} {\rm~GeV} \, ,
\\ \nonumber && \Gamma = 0.31 \pm 0.12 ^{+0.07}_{-0.08} {\rm~GeV} \, ;
\\ X(6600) &:& M = 6.62 \pm 0.03 ^{+0.02}_{-0.01} {\rm~GeV} \, ,
\\ \nonumber && \Gamma = 0.31 \pm 0.09 ^{+0.06}_{-0.11} {\rm~GeV} \, ;
\\ X(6900) &:& M = 6.87 \pm 0.03 ^{+0.06}_{-0.01} {\rm~GeV} \, ,
\\ \nonumber && \Gamma = 0.12 \pm 0.04 ^{+0.03}_{-0.01} {\rm~GeV} \, .
\end{eqnarray}
The ATLAS collaboration also investigated the $J/\psi \psi(2S)$ invariant mass spectrum. They reported the evidence for an enhancement at 6.9~GeV and a resonance at 7.2~GeV, whose masses and widths were measured to be~\cite{ATLAS}:
\begin{eqnarray}
X(6900) &:& M = 6.78 \pm 0.36 ^{+0.35}_{-0.54} {\rm~GeV} \, ,
\\ \nonumber && \Gamma = 0.39 \pm 11 ^{+0.11}_{-0.07} {\rm~GeV} \, ;
\\ X(7200) &:& M = 7.22 \pm 0.03 ^{+0.02}_{-0.03} {\rm~GeV} \, ,
\\ \nonumber && \Gamma = 0.10 {^{+0.13}_{-0.07}} {^{+0.06}_{-0.05}} {\rm~GeV} \, .
\end{eqnarray}

Actually, the fully-heavy tetraquark states were already studied by some theorists in the 1980s~\cite{Chao:1980dv,Iwasaki:1975pv,Ader:1981db,Heller:1985cb,Badalian:1985es,Zouzou:1986qh,Lloyd:2003yc,Barnea:2006sd,Berezhnoy:2011xn}, but there have  not been many relevant experiments from that time till now. Besides the above experiments~\cite{LHCb:2020bwg,CMS,ATLAS}, in 2017 the CMS collaboration found an excess in the $\Upsilon(1S)\mu^+\mu^-$ invariant mass spectrum near 18.5~GeV with a global significance of 3.6$\sigma$~\cite{Durgut:2018lmn,Yi:2018fxo}, and in 2019 the ANDY collaboration at RHIC reported an evidence of a significance peak at around 18.12 GeV~\cite{ANDY:2019bfn}. These structures are good candidates for the fully-bottom tetraquark states, and they also attracted some attention from the particle physics community~\cite{Anwar:2017toa,Esposito:2018cwh,Hughes:2017xie,Karliner:2016zzc,Wu:2016vtq,Richard:2017vry,Bai:2016int,Chen:2019dvd,Debastiani:2017msn,Wang:2019rdo,Liu:2019zuc,Deng:2020iqw,Richard:2018yrm,Wang:2017jtz,Wang:2018poa,Chiu:2007km}. Since these structures were not confirmed by some other experiments~\cite{LHCb:2018uwm,CMS:2020qwa}, they require more investigations crucially.

\begin{table}[t]
\begin{center}
\renewcommand{\arraystretch}{1.5}
\caption{Mass spectra of the fully-charm and fully-bottom tetraquark states, calculated in Ref.~\cite{Chen:2016jxd} through the QCD sum rule method.
\label{tab:mass}}
\begin{tabular}{c | c c c}
\hline\hline
~~$J^{PC}$~~ & ~~Currents~~ & ~~ $T_{c c\bar c \bar c}$ \mbox{(GeV)} ~~ & ~~ $T_{b b \bar b \bar b}$ \mbox{(GeV)} ~~
\\ \hline\hline
\multirow{2}{*}{$0^{++}$}      & $J^{0^{++}}_1$              &  $6.44\pm0.15$              & $18.45\pm0.15$
\\ \cline{2-4}
                               & $J^{0^{++}}_2$              &  $6.46\pm0.16$              & $18.46\pm0.14$
\\ \hline
$1^{+-}$                       & $J^{1^{+-}}_{3\alpha}$      &  $6.51\pm0.15$              & $18.54\pm0.15$
\\ \hline
$2^{++}$                       & $J^{2^{++}}_{4\alpha\beta}$ &  $6.51\pm0.15$              & $18.53\pm0.15$
\\ \hline\hline
\multirow{2}{*}{$0^{-+}$}      & $J^{0^{-+}}_5$              &  $6.84\pm0.18$              & $18.77\pm0.18$
\\ \cline{2-4}
                               & $J^{0^{-+}}_6$              &  $6.85\pm0.18$              & $18.79\pm0.18$
\\ \hline
$0^{--}$                       & $J^{0^{--}}_7$              &  $6.84\pm0.18$              & $18.77\pm0.18$
\\ \hline
\multirow{2}{*}{$1^{-+}$}      & $J^{1^{-+}}_{8\alpha}$      &  $6.84\pm0.18$              & $18.80\pm0.18$
\\ \cline{2-4}
                               & $J^{1^{-+}}_{9\alpha}$      &  $6.88\pm0.18$              & $18.83\pm0.18$
\\ \hline
\multirow{2}{*}{$1^{--}$}      & $J^{1^{--}}_{10\alpha}$     &  $6.84\pm0.18$              & $18.77\pm0.18$
\\ \cline{2-4}
                               & $J^{1^{--}}_{11\alpha}$     &  $6.83\pm0.18$              & $18.77\pm0.16$
\\ \hline\hline
\end{tabular}
\end{center}
\end{table}

In Ref.~\cite{Chen:2020xwe} we have studied the two-body decay behaviors of the fully-charm tetraquark states within the diquark-antidiquark picture, and calculated their relative branching ratios through the Fierz rearrangement of the Dirac and color indices. In this paper we shall further study the decay behaviors of the fully-bottom tetraquark states. As summarized in Table~\ref{tab:mass}, our previous QCD sum rule results suggest that the fully-charm tetraquark states lie above the di-charmonium thresholds, while some of the fully-bottom tetraquark states lie below the di-bottomonium thresholds~\cite{Chen:2016jxd}. Accordingly, in this paper we shall investigate not the fall-apart two-body decays, but the three-body decays of the fully-bottom tetraquark states. We shall study their decays into one bottomonium meson and one muon-antimuon pair, with the muon-antimuon pair produced by another intermediate vector bottomonium meson. This method has been applied in Refs.~\cite{Chen:2019wjd,Chen:2019eeq,Chen:2020pac,Chen:2020opr,Chen:2021erj} to investigate some other exotic hadrons, such as the $Z_c(3900)$ and the $P_c$ states, etc. Especially, our results obtained in Ref.~\cite{Chen:2019wjd} for the $Z_c(3900)$ are consistent with those obtained in Refs.~\cite{Dias:2013xfa,Agaev:2016dev,Esposito:2014hsa} using the QCD sum rule method and the non-relativistic effective field theory. Besides, a similar arrangement of the spin and color indices in the nonrelativistic case was applied in Refs.~\cite{Voloshin:2019aut,Voloshin:2013dpa,Maiani:2017kyi,Wang:2019spc,Xiao:2019spy,Cheng:2020nho} to study the decay properties of exotic hadrons, and our results obtained in Ref.~\cite{Chen:2020pac} for the $P_c$ states are consistent with those obtained in Ref.~\cite{Voloshin:2019aut} through the heavy quark spin symmetry.

This paper is organized as follows. In Sec.~\ref{sec:current} we construct the fully-bottom tetraquark currents within the diquark-antidiquark picture, and apply the Fierz rearrangement to transform them into the meson-meson currents. Based on the obtained Fierz identities, we study the decay behaviors of the fully-bottom tetraquark states in Sec.~\ref{sec:decay}, and calculate their relative branching ratios. The results are summarized and discussed in Sec.~\ref{sec:summary}.

\section{Currents and Fierz Identities}
\label{sec:current}

In Refs.~\cite{Chen:2016jxd,Chen:2020xwe} we have systematically constructed all the fully-heavy tetraquark currents without derivatives. We briefly summarize them in this section, which will be used to study the decay behaviors of the fully-bottom tetraquark states in the next section.

Besides, in Ref.~\cite{Su:2022eun} we have systematically constructed all the $P$-wave fully-strange tetraquark currents by explicitly adding the covariant derivative operator. Their corresponding fully-heavy tetraquark currents with derivatives can be similarly constructed. See Refs.~\cite{Chen:2020aos,Chen:2020uif} for more discussions. However, we shall not investigate them in the present study, since many relevant decay constants are not known yet. We also refer to Refs.~\cite{liu:2020eha,Liu:2020lpw} where the authors systematically classified the $S$-/$P$-wave fully-heavy/strange tetraquark states using the nonrelativistic quark model within the diquark-antidiquark picture.

\subsection{Currents of the positive parity}

There are altogether twelve fully-bottom tetraquark currents of the positive parity, four of which correspond to the $S$-wave fully-bottom tetraquark states within the diquark-antidiquark picture:
\begin{eqnarray}
J^{0^{++}}_1 &=& b_a^T C \gamma_5 b_b~\bar b_{a} \gamma_5 C \bar b_{b}^T \, ,
\label{def:current1}
\\ J^{0^{++}}_2 &=& b_a^T C \gamma_\mu b_b~\bar b_{a} \gamma^\mu C \bar b_{b}^T \, ,
\label{def:current2}
\\ J^{1^{+-}}_{3\alpha} &=& b_a^T C \gamma^\mu b_b~\bar b_{a} \sigma_{\alpha\mu} \gamma_5 C \bar b_{b}^T
\label{def:current3}
\\ \nonumber && ~~~~~~~~~~~~~~ - b_a^T C \sigma_{\alpha\mu} \gamma_5 b_b~\bar b_{a} \gamma^\mu C \bar b_{b}^T \, ,
\\ J^{2^{++}}_{4\alpha\beta} &=& \mathcal{P}_{\alpha\beta}^{\mu\nu}~b_a^T C \gamma_\mu b_b~\bar b_{a} \gamma_\nu C \bar b_{b}^T \, .
\label{def:current4}
\end{eqnarray}
In the above expressions $b_a$ is the bottom quark field with the color index $a$, and $\mathcal{P}_{\alpha\beta}^{\mu\nu}$ is the projection operator,
\begin{eqnarray}
\mathcal{P}^{\alpha\beta;\mu\nu} = g^{\alpha\mu} g^{\beta\nu} + g^{\alpha\nu} g^{\beta\mu} - {1\over2} g^{\alpha\beta} g^{\mu\nu}\, .
\end{eqnarray}
After applying the Fierz transformation, we obtain:
\begin{eqnarray}
\nonumber J^{0^{++}}_1 &=& -{1\over4} \xi_1^{0^{++}} -{1\over4} \xi_2^{0^{++}} -{1\over4} \xi_3^{0^{++}} -{1\over4} \xi_4^{0^{++}} +{1\over8} \xi_5^{0^{++}} \, ,
\\ \label{eq:tran1}
\\ J^{0^{++}}_2 &=& \xi_1^{0^{++}} - \xi_2^{0^{++}} + {1\over2} \xi_3^{0^{++}} - {1\over2} \xi_4^{0^{++}} \, ,
\label{eq:tran2}
\\ J^{1^{+-}}_{3\alpha} &=& {3i} \xi_{6\alpha}^{1^{+-}} - \xi_{7\alpha}^{1^{+-}} \, ,
\label{eq:tran3}
\\ J^{2^{++}}_{4\alpha\beta} &=& {1\over2} \xi_{8\alpha\beta}^{2^{++}} - {1\over2} \xi_{9\alpha\beta}^{2^{++}} + {1\over2} \xi_{10\alpha\beta}^{2^{++}} \, ,
\label{eq:tran4}
\end{eqnarray}
where
\begin{eqnarray}
   \nonumber \xi_1^{0^{++}} &=& \bar b_a b_a~\bar b_{b} b_b \, ,
\\ \nonumber \xi_2^{0^{++}} &=& \bar b_a \gamma_5 b_a~\bar b_{b} \gamma_5 b_b \, ,
\\ \nonumber \xi_3^{0^{++}} &=& \bar b_a \gamma_\mu b_a~\bar b_{b} \gamma^\mu b_b \, ,
\\ \nonumber \xi_4^{0^{++}} &=& \bar b_a \gamma_\mu \gamma_5 b_a~\bar b_{b} \gamma^\mu \gamma_5 b_b \, ,
\\ \xi_5^{0^{++}} &=& \bar b_a \sigma_{\mu\nu} b_a~\bar b_{b} \sigma^{\mu\nu} b_b \, ,
\\ \nonumber \xi_{6\alpha}^{1^{+-}} &=& \bar b_a \gamma_5 b_a~\bar b_{b} \gamma_\alpha b_b \, ,
\\ \nonumber \xi_{7\alpha}^{1^{+-}} &=& \bar b_a \gamma^\mu \gamma_5 b_a~\bar b_{b} \sigma_{\alpha\mu} b_b \, ,
\\ \nonumber \xi_{8\alpha\beta}^{2^{++}} &=& \mathcal{P}_{\alpha\beta}^{\mu\nu}~\bar b_a \gamma_\mu \gamma_5 b_a~\bar b_{b} \gamma_\nu \gamma_5 b_b \, ,
\\ \nonumber \xi_{9\alpha\beta}^{2^{++}} &=& \mathcal{P}_{\alpha\beta}^{\mu\nu}~\bar b_a \gamma_\mu b_a~\bar b_{b} \gamma_\nu b_b \, ,
\\ \nonumber \xi_{10\alpha\beta}^{2^{++}} &=& \mathcal{P}_{\alpha\beta}^{\mu\nu}~\bar b_a \sigma_{\mu\rho} b_a~\bar b_{b} \sigma_{\nu\rho} b_b \, .
\end{eqnarray}

\subsection{Currents of the negative parity}

There are altogether seven fully-bottom tetraquark currents of the negative parity:
\begin{eqnarray}
J^{0^{-+}}_5    &=& b^T_aCb_b ~ \bar{b}_a \gamma_5 C \bar{b}_b^T + b^T_aC\gamma_5b_b ~ \bar{b}_aC\bar{b}_b^T \, ,
\label{def:current5}
\\ J^{0^{-+}}_6 &=& b^T_a C \sigma_{\mu\nu} b_b ~ \bar{b}_a \sigma^{\mu\nu} \gamma_5 C \bar{b}^T_b \, ,
\label{def:current6}
\\ J^{0^{--}}_7 &=& b^T_a C b_b ~ \bar{b}_a\gamma_5C\bar{b}_b^T - b^T_aC\gamma_5b_b ~ \bar{b}_aC\bar{b}_b^T \, ,
\label{def:current7}
\\ \nonumber J^{1^{-+}}_{8\alpha} &=& b^T_aC\gamma_\alpha\gamma_5 b_b ~ \bar{b}_a\gamma_5C\bar{b}_b^T + b^T_aC\gamma_5b_b ~ \bar{b}_a\gamma_\alpha\gamma_5 C\bar{b}_b^T \, ,
\\ \label{def:current8}
\\ \nonumber J^{1^{-+}}_{9\alpha} &=& b^T_aC\sigma_{\alpha\mu}b_b ~ \bar{b}_a\gamma^\mu C\bar{b}^T_b+ b^T_aC\gamma^\mu b_b ~ \bar{b}_a\sigma_{\alpha\mu}C\bar{b}^T_b \, ,
\\ \label{def:current9}
\\ \nonumber J^{1^{--}}_{10\alpha} &=& b^T_aC\gamma_\alpha\gamma_5 b_b ~ \bar{b}_a\gamma_5C\bar{b}_b^T - b^T_aC\gamma_5b_b ~ \bar{b}_a\gamma_\alpha\gamma_5 C\bar{b}_b^T \, ,
\\ \label{def:current10}
\\ \nonumber J^{1^{--}}_{11\alpha} &=& b^T_aC\sigma_{\alpha\mu}b_b ~ \bar{b}_a\gamma^\mu C\bar{b}^T_b - b^T_aC\gamma^\mu b_b ~ \bar{b}_a\sigma_{\alpha\mu}C\bar{b}^T_b \, .
\\ \label{def:current11}
\end{eqnarray}
After applying the Fierz transformation, we obtain:
\begin{eqnarray}
J^{0^{-+}}_5 &=& -\xi_{11}^{0^{-+}} + {1\over4}\xi_{12}^{0^{-+}} \, ,
\label{eq:tran5}
\\ J^{0^{-+}}_6 &=& 6\xi_{11}^{0^{-+}} - {1\over2}\xi_{12}^{0^{-+}} \, ,
\label{eq:tran6}
\\ J^{0^{--}}_7 &=& - \xi_{13}^{0^{--}} \, ,
\label{eq:tran7}
\\ J^{1^{-+}}_{8\alpha} &=& - \xi_{14\alpha}^{1^{-+}} + i \xi_{15\alpha}^{1^{-+}} \, ,
\label{eq:tran8}
\\ J^{1^{-+}}_{9\alpha} &=& -3i \xi_{14\alpha}^{1^{-+}} + \xi_{15\alpha}^{1^{-+}} \, ,
\label{eq:tran9}
\\ J^{1^{--}}_{10\alpha} &=& \xi_{16\alpha}^{1^{--}} - i \xi_{17\alpha}^{1^{--}} \, ,
\label{eq:tran10}
\\ J^{1^{--}}_{11\alpha} &=& -3i \xi_{16\alpha}^{1^{--}} + \xi_{17\alpha}^{1^{--}} \, ,
\label{eq:tran11}
\end{eqnarray}
where
\begin{eqnarray}
\nonumber \xi_{11}^{0^{-+}} &=& \bar b_a b_a~\bar b_{b} \gamma_5 b_b \, ,
\\ \nonumber \xi_{12}^{0^{-+}} &=& \bar b_a \sigma_{\mu\nu} b_a~\bar b_{b} \sigma^{\mu\nu} \gamma_5 b_b \, ,
\\ \nonumber \xi_{13}^{0^{--}} &=& \bar b_a \gamma_\mu b_a~\bar b_{b} \gamma^\mu \gamma_5 b_b \, ,
\\ \xi_{14\alpha}^{1^{-+}} &=& \bar b_a \gamma_5 b_a~\bar b_{b} \gamma_\alpha \gamma_5 b_b \, ,
\\ \nonumber \xi_{15\alpha}^{1^{-+}} &=& \bar b_a \gamma^\mu b_a~\bar b_{b} \sigma_{\alpha\mu} b_b \, ,
\\ \nonumber \xi_{16\alpha}^{1^{--}} &=& \bar b_a b_a~\bar b_{b} \gamma_\alpha b_b \, ,
\\ \nonumber \xi_{17\alpha}^{1^{--}} &=& \bar b_a \gamma^\mu \gamma_5 b_a~\bar b_{b} \sigma_{\alpha\mu} \gamma_5 b_b \, .
\end{eqnarray}

\section{Relative Branching Ratios}
\label{sec:decay}

In this section we study possible decay channels of the fully-bottom tetraquark states, and calculate their relative branching ratios. The same method has been applied in Ref.~\cite{Chen:2020xwe} to study the fully-charm tetraquark states. We separately investigate the $S$- and $P$-wave fully-bottom tetraquark states as follows:
\begin{itemize}

\item Accordingly to our previous QCD sum rule calculations summarized in Table~\ref{tab:mass}, we assume the masses of the $S$-wave fully-bottom tetraquark states to be about 18.5~GeV. This value is below the $\eta_b(1S) \eta_b(1S)/\eta_b(1S) \Upsilon(1S)/\Upsilon(1S)\Upsilon(1S)$ thresholds, so the $S$-wave fully-bottom tetraquark states can not fall-apart decay into these two-body channels. Instead, they can decay into one bottomonium meson and one intermediate $\Upsilon(1S)/\Upsilon(2S)/\cdots$ meson, with the intermediate meson annihilating to be a photon and then transferring into a muon-antimuon pair. We depict this decay process in Fig.~\ref{fig:diagram}.

%
%%%%%%%%%%%%%%%%%%%%%%%%%%%%%%%%%%%%%%%%%%%%%%%%%%%%%%%%%%%%%%%%%%%%%%%%%%%%%%
%---------figure current 1
\begin{figure}[hbt]
\begin{center}
\includegraphics[width=0.45\textwidth]{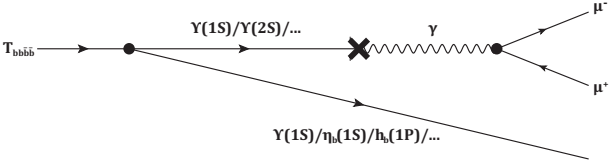}
\caption{Decay mechanism of a fully-bottom tetraquark state into one bottomonium meson and one intermediate $\Upsilon(1S)/\Upsilon(2S)/\cdots$ meson, with the intermediate $\Upsilon(1S)/\Upsilon(2S)/\cdots$ meson annihilating to be a photon and then transferring into a muon-antimuon pair.}
\label{fig:diagram}
\end{center}
\end{figure}
%%%%%%%%%%%%%%%%%%%%%%%%%%%%%%%%%%%%%%%%%%%%%%%%%%%%%%%%%%%%%%%%%%%%%%%%%%%%%%
%

\item Accordingly to our previous QCD sum rule calculations summarized in Table~\ref{tab:mass}, we assume the masses of the $P$-wave fully-bottom tetraquark states to be about 18.8~GeV. This value is below the $\eta_b(1S) \Upsilon(1S)/\Upsilon(1S) \Upsilon(1S)$ thresholds, so the $P$-wave fully-bottom tetraquark states can not fall-apart decay into these two-body channels. This value is above the $\eta_b(1S) \eta_b(1S)$ threshold, but the $P$-wave fully-bottom tetraquark states of $J^{PC} = 0^{-\pm}/1^{-\pm}$ can not decay into this two-body channel neither, due to either the $C$-parity conservation or the Bose-Einstein statistics. Accordingly, we shall also investigate the three-body decay process depicted in Fig.~\ref{fig:diagram}.

\end{itemize}

\begin{table*}[hbt]
\begin{center}
\renewcommand{\arraystretch}{2}
\caption{Couplings of the charmonium/bottomonium operators to the charmonium/bottomonium states.}
\begin{tabular}{ c | c | c | c | c | c}
\hline\hline
~~~Operators~~~ & ~~~$J^{PC}$~~~ & ~~~Mesons~~~ & ~~~$J^{PC}$~~~ & ~~~~~~~~~~~~~~~~~~~~Couplings~~~~~~~~~~~~~~~~~~~~ & ~~~~~~~~~~Decay Constants~~~~~~~~~~
\\ \hline\hline
$I^{S} = \bar c c$                                 & $0^{++}$   & $\chi_{c0}(1P)$  & $0^{++}$  & $\langle 0 | I^S | \chi_{c0}(1P) \rangle = m_{\chi_{c0}(1P)} f_{\chi_{c0}(1P)}$     & $f_{\chi_{c0}(1P)} = 343$~MeV~\mbox{\cite{Veliev:2010gb}}
\\ \hline
\multirow{2}{*}{$I^{P} = \bar c i\gamma_5 c$}  & \multirow{2}{*}{$0^{-+}$} & $\eta_c(1S)$     & $0^{-+}$  & $\langle 0 | I^{P} | \eta_c(1S) \rangle = \lambda_{\eta_c(1S)}$  & $\lambda_{\eta_c(1S)} = {f_{\eta_c(1S)} m_{\eta_c(1S)}^2 \over 2 m_c}$
\\ \cline{3-6}
                                                       && $\eta_c(2S)$     & $0^{-+}$  & $\langle 0 | I^{P} | \eta_c(2S) \rangle = \lambda_{\eta_c(2S)}$ & $\lambda_{\eta_c(2S)} = \lambda_{\eta_c(1S)} \times {f_{\psi(2S)} \over f_{J/\psi}}$
\\ \hline
\multirow{2}{*}{$I^{V}_\mu = \bar c \gamma_\mu c$} & \multirow{2}{*}{$1^{--}$} & $J/\psi$  & $1^{--}$  & $\langle 0 | I^{V}_\mu | J/\psi \rangle = m_{J/\psi} f_{J/\psi} \epsilon_\mu$  & $f_{J/\psi} = 418$~MeV~\mbox{\cite{Becirevic:2013bsa}}
\\ \cline{3-6}
                                                                && $\psi(2S)$  & $1^{--}$  & $\langle 0 | I^{V}_\mu | \psi(2S) \rangle = m_{\psi(2S)} f_{\psi(2S)} \epsilon_\mu$  & $f_{\psi(2S)} = 294.5$~MeV~\mbox{\cite{MaiordeSousa:2012vv}}
\\ \hline
\multirow{3}{*}{$I^{A}_\mu = \bar c \gamma_\mu \gamma_5 c$} & \multirow{3}{*}{$1^{++}$} & $\eta_c(1S)$ & $0^{-+}$ & $\langle 0 | I^{A}_\mu |\eta_c(1S) \rangle = i p_\mu f_{\eta_c(1S)}$ & $f_{\eta_c(1S)} = 387$~MeV~\mbox{\cite{Becirevic:2013bsa}}
\\ \cline{3-6}
&& $\eta_c(2S)$ & $0^{-+}$ & $\langle 0 | I^{A}_\mu | \eta_c(2S) \rangle = i p_\mu f_{\eta_c(2S)}$       & $f_{\eta_c(2S)} = f_{\eta_c(1S)} \times {f_{\psi(2S)} \over f_{J/\psi}}$
\\ \cline{3-6}
&&  $\chi_{c1}(1P)$   & $1^{++}$ &  $\langle 0 | I^{A}_\mu | \chi_{c1}(1P) \rangle = m_{\chi_{c1}(1P)} f_{\chi_{c1}(1P)} \epsilon_\mu $  &  $f_{\chi_{c1}(1P)} = 335$~MeV~\mbox{\cite{Novikov:1977dq}}
\\ \hline
\multirow{3}{*}{$I^{T}_{\mu\nu} = \bar c \sigma_{\mu\nu} c$} & \multirow{3}{*}{$1^{\pm-}$} & $J/\psi$ & $1^{--}$ & $\langle 0 | I^{T}_{\mu\nu} | J/\psi \rangle = i f^T_{J/\psi} (p_\mu\epsilon_\nu - p_\nu\epsilon_\mu) $  &  $f_{J/\psi}^T = 410$~MeV~\mbox{\cite{Becirevic:2013bsa}}
\\ \cline{3-6}
&& $\psi(2S)$ & $1^{--}$ & $\langle 0 | I^{T}_{\mu\nu} | \psi(2S) \rangle = i f^T_{\psi(2S)} (p_\mu\epsilon_\nu - p_\nu\epsilon_\mu) $  &  $f_{\psi(2S)}^T = f_{J/\psi}^T \times {f_{\psi(2S)} \over f_{J/\psi}}$
\\ \cline{3-6}
&&  $h_c(1P)$   & $1^{+-}$ &  $\langle 0 | I^{T}_{\mu\nu} | h_c(1P) \rangle = i f^T_{h_c(1P)} \epsilon_{\mu\nu\alpha\beta} \epsilon^\alpha p^\beta $  &  $f_{h_c(1P)}^T = 235$~MeV~\mbox{\cite{Becirevic:2013bsa}}
%%%%%%%%%%%%%%%%%%%%%%%%%%%%%%%%%%%%%%%%%%%%%%%%%%%%%%%%%%%%%%%%%%%%%%%%%%%
%%%%%%%%%%%%%%%%%%%%%%%%%%%%%%%%%%%%%%%%%%%%%%%%%%%%%%%%%%%%%%%%%%%%%%%%%%%
\\ \hline\hline
$J^{S} = \bar b b$                                 & $0^{++}$   & $\chi_{b0}(1P)$  & $0^{++}$  & $\langle 0 | J^S | \chi_{b0}(1P) \rangle = m_{\chi_{b0}(1P)} f_{\chi_{b0}(1P)}$     & $f_{\chi_{b0}(1P)} = 175$~MeV~\mbox{\cite{Veliev:2010gb}}
\\ \hline
\multirow{2}{*}{$J^{P} = \bar b i\gamma_5 b$}  & \multirow{2}{*}{$0^{-+}$} & $\eta_b(1S)$     & $0^{-+}$  & $\langle 0 | J^{P} | \eta_b(1S) \rangle = \lambda_{\eta_b(1S)}$  & $\lambda_{\eta_b(1S)} = {f_{\eta_b(1S)} m_{\eta_b(1S)}^2 \over 2 m_b}$
\\ \cline{3-6}
&& $\eta_b(2S)$     & $0^{-+}$  & $\langle 0 | J^{P} | \eta_b(2S) \rangle = \lambda_{\eta_b(2S)}$ & $\lambda_{\eta_b(2S)} = \lambda_{\eta_b(1S)} \times {f_{\Upsilon(2S)} \over f_{\Upsilon(1S)}}$
\\ \hline
\multirow{2}{*}{$J^{V}_\mu = \bar b \gamma_\mu b$} & \multirow{2}{*}{$1^{--}$} & $\Upsilon(1S)$  & $1^{--}$  & $\langle 0 | J^{V}_\mu | \Upsilon(1S) \rangle = m_{\Upsilon(1S)} f_{\Upsilon(1S)} \epsilon_\mu$  & $f_{\Upsilon(1S)} = 715$~MeV~\mbox{\cite{MaiordeSousa:2012vv}}
\\ \cline{3-6}
&& $\Upsilon(2S)$  & $1^{--}$  & $\langle 0 | J^{V}_\mu | \Upsilon(2S) \rangle = m_{\Upsilon(2S)} f_{\Upsilon(2S)} \epsilon_\mu$  & $f_{\Upsilon(2S)} = 497.5$~MeV~\mbox{\cite{MaiordeSousa:2012vv}}
\\ \hline
\multirow{3}{*}{$J^{A}_\mu = \bar b \gamma_\mu \gamma_5 b$} & \multirow{3}{*}{$1^{++}$} & $\eta_b(1S)$ & $0^{-+}$ & $\langle 0 | J^{A}_\mu |\eta_b(1S) \rangle = i p_\mu f_{\eta_b(1S)}$ & $f_{\eta_b(1S)} = 801$~MeV~\mbox{\cite{Chiu:2007km}}
\\ \cline{3-6}
&& $\eta_b(2S)$ & $0^{-+}$ & $\langle 0 | J^{A}_\mu | \eta_b(2S) \rangle = i p_\mu f_{\eta_b(2S)}$       & $f_{\eta_b(2S)} = f_{\eta_b(1S)} \times {f_{\Upsilon(2S)} \over f_{\Upsilon(1S)}}$
\\ \cline{3-6}
&&  $\chi_{b1}(1P)$   & $1^{++}$ &  $\langle 0 | J^{A}_\mu | \chi_{b1}(1P) \rangle = m_{\chi_{b1}(1P)} f_{\chi_{b1}(1P)} \epsilon_\mu $  &  $f_{\chi_{b1}(1P)} = f_{\chi_{b0}(1P)} \times {f_{\chi_{c1}(1P)} \over f_{\chi_{c0}(1P)}}$
\\ \hline
\multirow{3}{*}{$J^{T}_{\mu\nu} = \bar b \sigma_{\mu\nu} b$} & \multirow{3}{*}{$1^{\pm-}$} & $\Upsilon(1S)$ & $1^{--}$ & $\langle 0 | J^{T}_{\mu\nu} | \Upsilon(1S) \rangle = i f^T_{\Upsilon(1S)} (p_\mu\epsilon_\nu - p_\nu\epsilon_\mu) $  &  $f_{\Upsilon(1S)}^T = f^T_{J/\psi} \times {f_{\Upsilon(1S)} \over f_{J/\psi}}$
\\ \cline{3-6}
&& $\Upsilon(2S)$ & $1^{--}$ & $\langle 0 | J^{T}_{\mu\nu} | \Upsilon(2S) \rangle = i f^T_{\Upsilon(2S)} (p_\mu\epsilon_\nu - p_\nu\epsilon_\mu) $  &  $f_{\Upsilon(2S)}^T = f^T_{\Upsilon(1S)} \times {f_{\Upsilon(2S)} \over f_{\Upsilon(1S)}}$
\\ \cline{3-6}
&&  $h_b(1P)$   & $1^{+-}$ &  $\langle 0 | J^{T}_{\mu\nu} | h_b(1P) \rangle = i f^T_{h_b(1P)} \epsilon_{\mu\nu\alpha\beta} \epsilon^\alpha p^\beta $  &  $f_{h_b(1P)}^T = f_{h_c(1P)}^T \times {f_{\Upsilon(1S)} \over f_{J/\psi}}$
\\ \hline\hline
\end{tabular}
\label{tab:coupling}
\end{center}
\end{table*}

As an example, we apply the Fierz rearrangement given in Eq.~(\ref{eq:tran1}) to investigate the decay properties of the fully-bottom tetraquark state of $J^{PC} = 0^{++}$ corresponding to the current $J^{0^{++}}_1$ defined in Eq.~(\ref{def:current1}). We denote this state as $|X_1; 0^{++}\rangle$, and assume the coupling to be
\begin{equation}
\langle 0 | J^{0^{++}}_1 | X_1 ; 0^{++} \rangle = f_{X_1} \, ,
\end{equation}
with $f_{X_1}$ the decay constant. As summarized in Table~\ref{tab:mass}, its mass has been calculated in Ref.~\cite{Chen:2020xwe} through the QCD sum rule method to be $18.45\pm0.15$~GeV.

%
%%%%%%%%%%%%%%%%%%%%%%%%%%%%%%%%%%%%%%%%%%%%%%%%%%%%%%%%%%%%%%%%%%%%%%%%%%%%%%
%---------figure current 1
\begin{figure}[hbt]
\begin{center}
\includegraphics[width=0.25\textwidth]{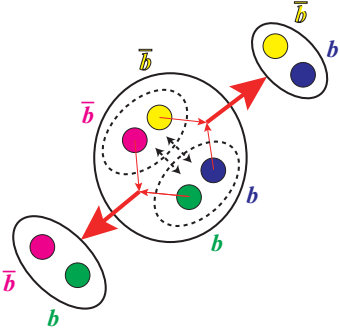}
\caption{The fall-apart decay of a compact diquark-antidiquark $[b b] [\bar b \bar b]$ state into two bottomonium states. Quarks are shown in red/green/blue color and antiquarks are shown in cyan/magenta/yellow color.}
\label{fig:decay}
\end{center}
\end{figure}
%%%%%%%%%%%%%%%%%%%%%%%%%%%%%%%%%%%%%%%%%%%%%%%%%%%%%%%%%%%%%%%%%%%%%%%%%%%%%%
%
As depicted in Fig.~\ref{fig:decay}, when the magenta $\bar b$ antiquark and the green $b$ quark meet each other, and the yellow $\bar b$ antiquark and the blue $b$ quark meet each other at the same time, $|X_1; 0^{++}\rangle$ can decay into two bottomonium states:
\begin{eqnarray}
&& \left[ b_a(x) b_b(x) \right] ~ \left[ \bar b_a(x) \bar b_b(x) \right]
\\[1mm] \nonumber &\xrightarrow{~\rm Fierz~}& \left[ \bar b_a(x) b_a(x) \right] ~ \left[ \bar b_b(x) b_b(x) \right]
\\[1mm] \nonumber &\xrightarrow{~\rm decay~}& \left[ \bar b_a(y) b_a(y) \right] ~ \left[ \bar b_b(z) b_b(z) \right] \, .
\end{eqnarray}
This decay process can be described by the Fierz rearrangement given in Eq.~(\ref{eq:tran1}), {\it i.e.},
\begin{eqnarray}
J^{0^{++}}_1 &=& b_a^T C \gamma_5 b_b~\bar b_{a} \gamma_5 C \bar b_{b}^T
\label{eq:example}
\\ \nonumber &\rightarrow& - {1\over4} \bar b_a b_a~\bar b_{b} b_b - {1\over4} \bar b_a \gamma_5 b_a~\bar b_{b} \gamma_5 b_b
\\ \nonumber && - {1\over4} \bar b_a \gamma_\mu \gamma_5 b_a~\bar b_{b} \gamma^\mu \gamma_5 b_b - {1\over4} \bar b_a \gamma_\mu b_a~\bar b_{b} \gamma^\mu b_b
\\ \nonumber && + {1\over8} \bar b_a \sigma_{\mu\nu} b_a~\bar b_{b} \sigma^{\mu\nu} b_b \, .
\end{eqnarray}
In principle, we need the decay constant $f_{X_1}$ as an input to calculate the partial decay widths, but it is not necessary any more if we only calculate the relative branching ratios. Moreover, because the couplings of meson operators to meson states have been well studied in the literature, but the couplings of tetraquark currents to tetraquark states have not, the decay constant $f_{X_1}$ is not so well determined compared to the bottomonium decay constants listed in Table~\ref{tab:coupling}. Therefore, we can calculate the relative branching ratios more reliably than the partial decay widths.

To do this we apply Eq.~(\ref{eq:example}) to derive the couplings of the current $J^{0^{++}}_1$ to both the $\Upsilon(1S)\Upsilon(1S)$ and $\Upsilon(1S) h_b(1P)$ channels to be:
\begin{eqnarray}
&& \langle 0 | J^{0^{++}}_1 | \Upsilon(p_1,\epsilon_1)~\Upsilon(p_2,\epsilon_2) \rangle
\\ \nonumber &=& \epsilon_1^\mu \epsilon_2^\nu ~ \Big( - {1\over2} m_{\Upsilon}^2 f_{\Upsilon}^2 g_{\mu\nu}
\\ \nonumber && ~~~~~ - {1\over2} (f_{\Upsilon}^{T})^2 p_1 \cdot p_2 g_{\mu\nu} + {1\over2} (f_{\Upsilon}^{T})^2 p_{1\nu} p_{2\mu} \Big) \, ,
\\
&& \langle 0 | J^{0^{++}}_1 | \Upsilon(p_1,\epsilon_1)~h_b(p_2,\epsilon_2) \rangle
\\ \nonumber &=& - {1\over2} \times \epsilon_1^\mu \epsilon_2^\nu ~ f_{\Upsilon}^{T} f_{h_b}^{T} \epsilon_{\mu\nu\rho\sigma} p_{1}^{\rho} p_{2}^{\sigma} \, ,
\end{eqnarray}
from which we further extract the couplings of $|X_1 ; 0^{++} \rangle$ to both the $\Upsilon(1S)\Upsilon(1S)$ and $\Upsilon(1S) h_b(1P)$ channels to be:
\begin{eqnarray}
&& \langle X_1(p); 0^{++} | \Upsilon(p_1,\epsilon_1)~\Upsilon(p_2,\epsilon_2) \rangle
\label{eq:X1Upsilon}
\\ \nonumber &=& c \times \epsilon_1^\mu \epsilon_2^\nu ~ \Big( - {1\over2} m_{\Upsilon}^2 f_{\Upsilon}^2 g_{\mu\nu}
\\ \nonumber && ~~~~~ - {1\over2} (f_{\Upsilon}^{T})^2 p_1 \cdot p_2 g_{\mu\nu} + {1\over2} (f_{\Upsilon}^{T})^2 p_{1\nu} p_{2\mu} \Big) \, ,
\\
&& \langle X_1(p); 0^{++} | \Upsilon(p_1,\epsilon_1)~h_b(p_2,\epsilon_2) \rangle
\\ \nonumber &=& - {c\over2} \times \epsilon_1^\mu \epsilon_2^\nu ~ f_{\Upsilon}^{T} f_{h_b}^{T} \epsilon_{\mu\nu\rho\sigma} p_{1}^{\rho} p_{2}^{\sigma} \, .
\end{eqnarray}
The decay constants $f_{\Upsilon} \equiv f_{\Upsilon(1S)}$, $f_{\Upsilon}^{T} \equiv f_{\Upsilon(1S)}^{T}$, and $f_{h_b}^{T} \equiv f_{h_b(1P)}^{T}$ of the $\Upsilon(1S)$ and $h_b(1P)$ mesons are given in Table~\ref{tab:coupling}. The overall factor $c$ is related to the decay constant $f_{X_1}$, which will be eliminated when calculating the relative branching ratios.

Based on Eq.~(\ref{eq:X1Upsilon}), we can write the decay amplitude of the three-body decay process $|X_1 ; 0^{++} \rangle \rightarrow \Upsilon(1S)\Upsilon(1S) \rightarrow \Upsilon(1S) \mu^+ \mu^-$ as
\begin{eqnarray}
&& \mathcal{M} \Big( X_1(p) \to \Upsilon(p_1,\epsilon_1) \Upsilon(q,\epsilon_2)
\label{eq:amplitude}
\\ \nonumber && ~~~~~~~~~~~~~ \rightarrow \Upsilon(p_1,\epsilon_1) \mu^-(p_2) \mu^+(p_3) \Big)
\\ \nonumber &=& c c^\prime e \times \epsilon_1^\mu ~ \Big( - {1\over2} m_{\Upsilon}^2 f_{\Upsilon}^2 g_{\mu\nu}
\\ \nonumber && ~~~~~ - {1\over2} (f_{\Upsilon}^{T})^2 p_1 \cdot q g_{\mu\nu} + {1\over2} (f_{\Upsilon}^{T})^2 p_{1\nu} q_{\mu} \Big)
\\ \nonumber && \times~{ \bar u(p_2) \gamma_\alpha v(p_3) \over q^2 \left( q^2 - m_{\Upsilon}^2 + {\rm i} m_{\Upsilon} \Gamma_{\Upsilon} \right) } ~ \left( g^{\alpha\nu} - {q^{\alpha} q^{\nu} \over m_{\Upsilon}^2} \right) \, ,
\end{eqnarray}
where $u(p_2)$ and $v(p_3)$ are the Dirac spinors of the $\mu^-$ and $\mu^+$, respectively. The overall factor $c^\prime$ is related to the coupling of $\Upsilon(1S)$ to the photon, which will also be eliminated when calculating the relative branching ratios.

We use Eq.~(\ref{eq:amplitude}) to further evaluate the partial decay width to be:
\begin{eqnarray}
&& \Gamma \left( X_1(p) \to \Upsilon(p_1,\epsilon_1) \mu^+(p_2) \mu^-(p_3) \right)
\\ \nonumber &=& {1\over(2\pi)^3}{ c^2 c^{\prime2} e^2 \over 32 m_{X_1}^3} \int {\rm d}m_{12}^2 {\rm d}m_{23}^2
\\ \nonumber && \times \left|{1 \over q^2 - m_{\Upsilon}^2 + {\rm i} m_{\Upsilon} \Gamma_{\Upsilon} }\right|^2 \left|{1 \over q^2 }\right|^2
\\ \nonumber && \times {\rm Tr}[(p\!\!\!\slash_2+m_{\mu^-}) \gamma_\alpha (p\!\!\!\slash_3-m_{\mu^+}) \gamma_{\alpha^\prime}]
\\ \nonumber && \times \Big( - {1\over2} m_{\Upsilon}^2 f_{\Upsilon}^2 g_{\mu\nu}
\\ \nonumber && ~~~~~ - {1\over2} (f_{\Upsilon}^{T})^2 p_1 \cdot q g_{\mu\nu} + {1\over2} (f_{\Upsilon}^{T})^2 p_{1\nu} q_{\mu} \Big)
\\ \nonumber && \times \Big( - {1\over2} m_{\Upsilon}^2 f_{\Upsilon}^2 g_{\mu^\prime\nu^\prime}
\\ \nonumber && ~~~~~ - {1\over2} (f_{\Upsilon}^{T})^2 p_1 \cdot q g_{\mu^\prime\nu^\prime} + {1\over2} (f_{\Upsilon}^{T})^2 p_{1\nu^\prime} q_{\mu^\prime} \Big)
\\ \nonumber && \times \left( g^{\alpha\nu} - {q^{\alpha} q^{\nu} \over m_{\Upsilon}^2} \right) ~ \left( g^{\alpha^\prime\nu^\prime} - {q^{\alpha^\prime} q^{\nu^\prime} \over m_{\Upsilon}^2} \right)
\\ \nonumber && \times \left( g^{\mu\mu^\prime} - {p_1^{\mu} p_1^{\mu^\prime} \over m_{\Upsilon}^2} \right) \, .
\end{eqnarray}

Similarly, we study the three-body decay process $|X_1 ; 0^{++} \rangle \rightarrow h_b(1P)\Upsilon(1S) \rightarrow h_b(1P) \mu^+ \mu^-$ and calculate its partial decay width. After eliminating the overall factors $c$ and $c^\prime$, we obtain:
\begin{equation}
{\mathcal{B}(| X_1 ; 0^{++} \rangle \rightarrow h_b(1P)\Upsilon(1S) \rightarrow h_b(1P) \mu^+ \mu^-) \over \mathcal{B}(| X_1 ; 0^{++} \rangle \rightarrow \Upsilon(1S)\Upsilon(1S) \rightarrow \Upsilon(1S) \mu^+ \mu^-)} = 0.002 \, .
\end{equation}
The above procedures are applied to investigate the process with the intermediate $\Upsilon(1S)$ meson. We can apply the same procedures to investigate the process with the intermediate $\Upsilon(2S)/\Upsilon(3S)/\cdots$ mesons, and the obtained results are approximately the same, while we do not consider other intermediate bottomonium mesons in the present study, such as the $\Upsilon(1D)$ meson, etc. Assuming that the $|X_1 ; 0^{++} \rangle \rightarrow \Upsilon(1S) \mu^+ \mu^-$ and $|X_1 ; 0^{++} \rangle \rightarrow h_b(1P) \mu^+ \mu^-$ decays are dominated by these processes, we finally obtain
\begin{equation}
{\mathcal{B}(| X_1 ; 0^{++} \rangle \rightarrow h_b(1P) \mu^+ \mu^-) \over \mathcal{B}(| X_1 ; 0^{++} \rangle \rightarrow \Upsilon(1S) \mu^+ \mu^-)} \approx 0.002 \, .
\end{equation}
After considering several relevant channels, we obtain:
\begin{eqnarray}
\nonumber \mathcal{B}(| X_1 ; 0^{++} \rangle &\rightarrow& \Upsilon(1S)\mu\mu : \Upsilon(2S)\mu\mu : h_b(1P)\mu\mu \,)
\\ &\approx& ~~~~~1~~~~~\, : ~~~0.42~~~ : ~~~0.002  \, .
\end{eqnarray}

Similarly, we apply the above procedures to investigate the $S$- and $P$-wave fully-bottom tetraquark states $|X_{2\cdots11}; J^{PC}\rangle$ through the currents $J^{\cdots}_{2\cdots11}$. The obtained results are summarized in Table~\ref{tab:result1}, which we shall use to draw conclusions in the next section. It is interesting to notice that some relative branching ratios are significantly larger/smaller than the others, which is partly due to that these ratios are proportional to the square of the Fierz coefficients given in Eqs.~(\ref{eq:tran1}-\ref{eq:tran4}) and Eqs.~(\ref{eq:tran5}-\ref{eq:tran11}). For example, the relative branching ratio of $|X_{11}; 1^{--}\rangle$ decaying into the $\mu^+ \mu^- \chi_{b0}(1P)$ channel is significantly larger than that of the $\mu^+ \mu^- \eta_b(1S)$ channel:
\begin{equation}
{\mathcal{B}(| X_{11}; 1^{--} \rangle \rightarrow \chi_{b0}(1P) \mu^+ \mu^-) \over \mathcal{B}(| X_{11}; 1^{--} \rangle \rightarrow \eta_b(1S) \mu^+ \mu^-)} \approx 12 \, ,
\label{eq:br}
\end{equation}
where the factor contributed by the Fierz coefficients is 9. Besides, the relative branching ratios are also contributed by the decay constants as well as the kinematics. For example, the relative branching ratio of $|X_{8}; 1^{-+}\rangle$ decaying into the $\mu^+ \mu^- h_b(1P)$ channel is not far from that of the $\mu^+ \mu^- \Upsilon(1S)$ channel:
\begin{equation}
{\mathcal{B}(| X_{8}; 1^{-+} \rangle \rightarrow h_b(1P) \mu^+ \mu^-) \over \mathcal{B}(| X_{8}; 1^{-+} \rangle \rightarrow \Upsilon(1S) \mu^+ \mu^-)} \approx 0.27 \, .
\end{equation}
The Fierz coefficients do not contribute to this ratio, while the factors contributed by the decay constants and the kinematics are about 0.33 and 0.81, respectively.

In our previous study~\cite{Chen:2020xwe} we have studied the decays of the fully-charm tetraquark states into the $1S$ and $1P$ double-charmonium channels $J/\psi J/\psi$, $J/\psi \eta_c(1S)$, and $\eta_c(1S) \eta_c(1S)$, etc. In the present study we further take into account the $2S$ double-charmonium channels $J/\psi \psi(2S)$, $\eta_c(1S) \eta_c(2S)$, $J/\psi \eta_c(2S)$, and $\eta_c(1S) \psi(2S)$. The obtained results are summarized in Table~\ref{tab:result2}, which we shall also use to draw conclusions in the next section.

%================================================================================
%================================================================================
\section{Summary and discussions}
\label{sec:summary}
%================================================================================
%================================================================================

\begin{table*}[hbt]
\begin{center}
\renewcommand{\arraystretch}{1.5}
\caption{Relative branching ratios of the $S$- and $P$-wave fully-bottom tetraquark states $|X_{1\cdots11}; J^{PC}\rangle$ corresponding to the currents $J^{\cdots}_{1\cdots11}$. In the 3rd-5th columns we show the branching ratios relative to the $\mu^+ \mu^- \Upsilon(1S)$ channel, and in the 6th-9th columns we show the branching ratios relative to the $\mu^+ \mu^- \eta_b(1S)$ channel.}
\begin{tabular}{ c | c | c c c | c c c c }
\hline \hline
\multirow{2}{*}{~~$J^{PC}$~~} & \multirow{2}{*}{~~Current~~} & \multicolumn{7}{c}{Decay Channels}
\\ \cline{3-9}
 && \,$\mu^+ \mu^- \Upsilon(1S)$\, & \,$\mu^+ \mu^- \Upsilon(2S)$\, & \,$\mu^+ \mu^- h_b(1P)$\,
& \,$\mu^+ \mu^- \eta_b(1S)$\, & \,$\mu^+ \mu^- \eta_b(2S)$\, & \,$\mu^+ \mu^- \chi_{b0}(1P)$\, & \,$\mu^+ \mu^- \chi_{b1}(1P)$\,
\\ \hline \hline
\multirow{2}{*}{$0^{++}$} & $J^{0^{++}}_1$               & $1$  & $0.42$  & $0.002$ & --  & --     & --      & --
\\ \cline{2-9}
                          & $J^{0^{++}}_2$               & $1$  & $0.42$  & --      & --  & --     & --      & --
\\ \hline
$1^{+-}$                  & $J^{1^{+-}}_{3\alpha}$       & --   & --      & --      & $1$ & $0.42$ & --      & $1 \times 10^{-4}$
\\ \hline
$2^{++}$                  & $J^{2^{++}}_{4\alpha\beta}$  & $1$  & $0.42$  & $0.002$ & --  & --     & --      & --
\\ \hline \hline
\multirow{2}{*}{$0^{-+}$} & $J^{0^{-+}}_5$               & $1$  & $0.39$  & $0.090$ & --  & --     & --      & --
\\ \cline{2-9}
                          & $J^{0^{-+}}_6$               & $1$  & $0.39$  & $0.090$ & --  & --     & --      & --
\\ \hline
$0^{--}$                  & $J^{0^{--}}_7$               & --   & --      & --      & $1$ & $0.42$ & --      & $0.041$
\\ \hline
\multirow{2}{*}{$1^{-+}$} & $J^{1^{-+}}_{8\alpha}$       & $1$  & $0.43$  & $0.27$   & --  & --     & --      & --
\\ \cline{2-9}
                          & $J^{1^{-+}}_{9\alpha}$       & $1$  & $0.43$  & $0.27$   & --  & --     & --      & --
\\ \hline
\multirow{2}{*}{$1^{--}$} & $J^{1^{--}}_{10\alpha}$      & --   & --      & --      & $1$ & $0.38$ & $1.3$ & $0.070$
\\ \cline{2-9}
                          & $J^{1^{--}}_{11\alpha}$      & --   & --      & --      & $1$ & $0.38$ & $12$  & $0.070$
\\ \hline \hline
\end{tabular}
\label{tab:result1}
\end{center}
\end{table*}

\begin{table*}[hbt]
\begin{center}
\renewcommand{\arraystretch}{1.5}
\caption{Relative branching ratios of the $S$- and $P$-wave fully-charm tetraquark states, calculated through the fully-charm tetraquark currents $J^{\cdots}_{1\cdots11}|_{b/\bar b \to c/\bar c}$. In the 3rd-9th columns we show the branching ratios relative to the $J/\psi J/\psi$ channel, and in the 10th-15th columns we show the branching ratios relative to the $J/\psi \eta_c$ channel. The notations $\psi^\prime \equiv \psi(2S)$ and $\eta_c^\prime \equiv \eta_c(2S)$ are used here.}
\begin{tabular}{ c | c | c  c  c  c  c  c  c | c  c  c  c  c  c }
\hline \hline
\multirow{2}{*}{~~$J^{PC}$~~} & \multirow{2}{*}{~~Current~~} & \multicolumn{13}{c}{Decay Channels}
\\ \cline{3-15}
&& \,$J/\psi J/\psi$\, & \,$J/\psi \psi^\prime$\, & ~$\eta_c \eta_c$~ & \,$\eta_c \eta_c^\prime$\, & \,$J/\psi h_c$\, &
\,$\eta_c \chi_{c0}$\, & \,$\eta_c \chi_{c1}$\, & ~$J/\psi \eta_c$~ & ~$J/\psi \eta_c^\prime$~ & ~$\psi^\prime \eta_c$~ & \,$J/\psi \chi_{c0}$\, & \,$J/\psi \chi_{c1}$\, & ~$\eta_c h_c$~
\\ \hline \hline
\multirow{2}{*}{$0^{++}$} & $J^{0^{++}}_1$               & $1$ & --      & $0.45$  & -- & --     & --     & $2\times10^{-5}$ & --  & --      & --      & --     & --    & --
\\ \cline{2-15}
                          & $J^{0^{++}}_2$               & $1$ & --      & $4.1$   & -- & --     & --     & $9\times10^{-5}$ & --  & --      & --      & --     & --    & --
\\ \hline
$1^{+-}$                  & $J^{1^{+-}}_{3\alpha}$       & --  & --      & --      & -- & --     & --     & --               & $1$ & --      & --      & --     & --    & --
\\ \hline
$2^{++}$                  & $J^{2^{++}}_{4\alpha\beta}$  & $1$ & --      & $0.036$ & -- & --     & --     & $0.003$          & --  & --      & --      & --     & --    & --
\\ \hline \hline
\multirow{2}{*}{$0^{-+}$} & $J^{0^{-+}}_5$               & $1$ & $0.071$ & --      & -- & $0.21$ & $0.69$ & --               & --  & --      & --      & --     & --    & --
\\ \cline{2-15}
                          & $J^{0^{-+}}_6$               & $1$ & $0.071$ & --      & -- & $0.21$ & $6.2$  & --               & --  & --      & --      & --     & --    & --
\\ \hline
$0^{--}$                  & $J^{0^{--}}_7$               & --  & --      & --      & -- & --     & --     & --               & $1$ & $0.048$ & $0.078$ & --     & $1.4$ & --
\\ \hline
\multirow{2}{*}{$1^{-+}$} & $J^{1^{-+}}_{8\alpha}$       & $1$ & $0.071$ & --      & -- & $0.78$ & --     & $0.94$           & --  & --      & --      & --     & --    & --
\\ \cline{2-15}
                          & $J^{1^{-+}}_{9\alpha}$       & $1$ & $0.071$ & --      & -- & $0.78$ & --     & $8.4$            & --  & --      & --      & --     & --    & --
\\ \hline
\multirow{2}{*}{$1^{--}$} & $J^{1^{--}}_{10\alpha}$      & --  & --      & --      & -- & --     & --     & --               & $1$ & $0.048$ & $0.078$ & $0.79$ & $1.5$ & $0.43$
\\ \cline{2-15}
                          & $J^{1^{--}}_{11\alpha}$      & --  & --      & --      & -- & --     & --     & --               & $1$ & $0.048$ & $0.078$ & $7.1$  & $1.5$ & $0.43$
\\ \hline \hline
\end{tabular}
\label{tab:result2}
\end{center}
\end{table*}

In this paper we systematically study the decay behaviors of the fully-bottom and fully-charm tetraquark states through their corresponding interpolating currents without derivatives. We work within the diquark-antidiquark picture, and apply the Fierz rearrangement of the Dirac and color indices to transform the diquark-antidiquark currents into the meson-meson currents. The obtained Fierz identities are given in Eqs.~(\ref{eq:tran1}-\ref{eq:tran4}) and Eqs.~(\ref{eq:tran5}-\ref{eq:tran11}).

Based on these Fierz identities, we study the decay mechanism depicted in Fig.~\ref{fig:diagram}, where a fully-bottom tetraquark state decays into one bottomonium meson and one intermediate $\Upsilon(1S)/\Upsilon(2S)/\cdots$ meson, with the intermediate $\Upsilon(1S)/\Upsilon(2S)/\cdots$ meson annihilating to be a photon and then transferring into a muon-antimuon pair. We consider several possible decay channels and calculate their relative branching ratios. The obtained results are summarized in Table~\ref{tab:result1}, where the masses of the $S$- and $P$-wave fully-bottom tetraquark states are assumed to be 18.5~GeV and 18.8~GeV, respectively~\cite{Chen:2016jxd}. In the calculations we work within the naive factorization scheme, so the uncertainty of our results is significantly larger than the well-developed QCD factorization scheme (about 5\% when studying the weak and radiative decays of the conventional hadrons)~\cite{Beneke:1999br,Beneke:2000ry,Beneke:2001ev,Li:2020rcg}. However, we calculate the relative branching ratios after eliminating several ambiguous overall factors, such as the decay constant $f_{X_1}$ and the coupling of the $\Upsilon(1S)$ to the photon. This largely reduces our uncertainty, {\it e.g.}, we roughly estimate the uncertainty of Eq.~(\ref{eq:br}) to be
\begin{equation}
{\mathcal{B}(| X_{11}; 1^{--} \rangle \rightarrow \chi_{b0}(1P) \mu^+ \mu^-) \over \mathcal{B}(| X_{11}; 1^{--} \rangle \rightarrow \eta_b(1S) \mu^+ \mu^-)} \approx 12^{+24}_{-~8} \, ,
\end{equation}
based on our previous systematical QCD sum rule studies on the decay properties of the excited heavy baryons~\cite{Yang:2020zrh,Yang:2021lce}.

Our results suggest that the fully-bottom tetraquark states of $J^{PC} = 0^{++}/2^{++}/0^{-+}/1^{-+}$ can be searched for in the $\mu^+ \mu^- \Upsilon(1S)$ channel, and they can also be searched for in the $\mu^+ \mu^- \Upsilon(2S)$ channel, with the relative branching ratio $\mathcal{B}(X \to \mu\mu \Upsilon(2S)) / \mathcal{B}(X \to \mu \mu \Upsilon(1S)) \approx 0.4$. Our results also suggest that the fully-bottom tetraquark states of $J^{PC} = 1^{+-}/0^{--}/1^{--}$ can be searched for in the $\mu^+ \mu^- \eta_b(1S)$ channel, and they can also be searched for in the $\mu^+ \mu^- \eta_b(2S)$ channel, with the similar relative branching ratio $\mathcal{B}(X \to \mu\mu \eta_b(2S)) / \mathcal{B}(X \to \mu \mu \eta_b(1S)) \approx 0.4$. We propose to examine these decay channels to search for the fully-bottom tetraquark states in future CMS experiments.

In this paper we also update our previous study of Ref.~\cite{Chen:2020xwe} and reanalysis the fall-apart two-body decays of the fully-charm tetraquark states. The obtained results are summarized in Table~\ref{tab:result2}, where the masses of the $S$- and $P$-wave fully-charm tetraquark states are assumed to be 6.5~GeV and 6.9~GeV, respectively~\cite{Chen:2016jxd}. These states were used in Ref.~\cite{Chen:2020xwe} to explain the broad structure at around 6.2-6.8~GeV and the narrow structure at around 6.9~GeV observed by LHCb in the di-$J/\psi$ invariant mass spectrum~\cite{LHCb:2020bwg}. Based on the results of the present study, we calculate the $X(6900)$ decay into the $J/\psi \psi(2S)$ channel and obtain the relative branching ratio $\mathcal{B}(X \to J/\psi \psi(2S)) / \mathcal{B}(X \to J/\psi J/\psi) \approx 0.1$.

%
%=====================================================================================
%=====================================================================================
%=====================================================================================
\section*{Acknowledgments}
%=====================================================================================
%=====================================================================================
%=====================================================================================
%

We thank Xiang Liu and Shi-Lin Zhu for helpful discussion.
This project is supported by
the National Natural Science Foundation of China under Grants No.~12075019 and No.~12175318,
the Jiangsu Provincial Double-Innovation Program under Grant No.~JSSCRC2021488,
the Natural Science Foundation of Guangdong Province of China under Grant No.~2022A1515011922,
and
the Fundamental Research Funds for the Central Universities.

%
%%%%%%%%%%%%%%%%%%%%%%%%%%%%%%%%%%%%%%%%%%%%%%%%%%%%%%%%%%%%%%%%%%%%%%%%%%%%%%


\begin{thebibliography}{99}

%\cite{LHCb:2020bwg}
\bibitem{LHCb:2020bwg}
R.~Aaij \textit{et al.} [LHCb Collaboration],
{\it Observation of structure in the $J /\psi$-pair mass spectrum},
\href{http://dx.doi.org/10.1016/j.scib.2020.08.032}{Sci. Bull. \textbf{65} (2020) no.23, 1983-1993}.

%\cite{liu:2020eha}
\bibitem{liu:2020eha}
M.~S.~Liu, F.~X.~Liu, X.~H.~Zhong and Q.~Zhao,
{\it Full-heavy tetraquark states and their evidences in the LHCb di-$J/\psi$ spectrum},
\href{http://arxiv.org/abs/arXiv:2006.11952}{arXiv:2006.11952 [hep-ph]}.
%40 citations counted in INSPIRE as of 12 May 2022

%\cite{Tiwari:2021tmz}
\bibitem{Tiwari:2021tmz}
R.~Tiwari, D.~P.~Rathaud and A.~K.~Rai,
{\it Spectroscopy of all charm tetraquark states},
\href{http://arxiv.org/abs/arXiv:2108.04017}{arXiv:2108.04017 [hep-ph]}.
%6 citations counted in INSPIRE as of 12 May 2022

%\cite{Lu:2020cns}
\bibitem{Lu:2020cns}
Q.~F.~L\"u, D.~Y.~Chen and Y.~B.~Dong,
{\it Masses of fully heavy tetraquarks $QQ {\bar{Q}} {\bar{Q}}$ in an extended relativized quark model},
\href{http://dx.doi.org/10.1140/epjc/s10052-020-08454-1}{Eur. Phys. J. C \textbf{80} (2020) no.9, 871}.
%doi:10.1140/epjc/s10052-020-08454-1
%[arXiv:2006.14445 [hep-ph]].
%59 citations counted in INSPIRE as of 12 May 2022

%\cite{Faustov:2020qfm}
\bibitem{Faustov:2020qfm}
R.~N.~Faustov, V.~O.~Galkin and E.~M.~Savchenko,
{\it Masses of the $QQ\bar Q\bar Q$ tetraquarks in the relativistic diquark--antidiquark picture},
\href{http://dx.doi.org/10.1103/PhysRevD.102.114030}{Phys. Rev. D \textbf{102} (2020) no.11, 114030}.
%[arXiv:2009.13237 [hep-ph]].
%27 citations counted in INSPIRE as of 12 May 2022

%\cite{Zhang:2020xtb}
\bibitem{Zhang:2020xtb}
J.~R.~Zhang,
{\it $0^{+}$ fully-charmed tetraquark states},
\href{http://dx.doi.org/10.1103/PhysRevD.103.014018}{Phys. Rev. D \textbf{103} (2021) no.1, 014018}.
%[arXiv:2010.07719 [hep-ph]].
%27 citations counted in INSPIRE as of 12 May 2022

%\cite{Li:2021ygk}
\bibitem{Li:2021ygk}
Q.~Li, C.~H.~Chang, G.~L.~Wang and T.~Wang,
{\it Mass spectra and wave functions of ${T}_{QQ\bar{Q}\bar{Q}}$ tetraquarks},
\href{http://dx.doi.org/10.1103/PhysRevD.104.014018}{Phys. Rev. D \textbf{104} (2021) no.1, 014018}.
%[arXiv:2104.12372 [hep-ph]].
%13 citations counted in INSPIRE as of 12 May 2022

%\cite{Bedolla:2019zwg}
\bibitem{Bedolla:2019zwg}
M.~A.~Bedolla, J.~Ferretti, C.~D.~Roberts and E.~Santopinto,
{\it Spectrum of fully-heavy tetraquarks from a diquark+antidiquark perspective},
\href{http://dx.doi.org/10.1140/epjc/s10052-020-08579-3}{Eur. Phys. J. C \textbf{80} (2020) no.11, 1004}.
%[arXiv:1911.00960 [hep-ph]].
%67 citations counted in INSPIRE as of 12 May 2022

%\cite{Weng:2020jao}
\bibitem{Weng:2020jao}
X.~Z.~Weng, X.~L.~Chen, W.~Z.~Deng and S.~L.~Zhu,
{\it Systematics of fully heavy tetraquarks},
\href{http://dx.doi.org/10.1103/PhysRevD.103.034001}{Phys. Rev. D \textbf{103} (2021) no.3, 034001}.
%[arXiv:2010.05163 [hep-ph]].
%34 citations counted in INSPIRE as of 12 May 2022

%\cite{Liu:2021rtn}
\bibitem{Liu:2021rtn}
F.~X.~Liu, M.~S.~Liu, X.~H.~Zhong and Q.~Zhao,
{\it Higher mass spectra of the fully-charmed and fully-bottom tetraquarks},
\href{http://dx.doi.org/10.1103/PhysRevD.104.116029}{Phys. Rev. D \textbf{104} (2021) no.11, 1160291}.
%[arXiv:2110.09052 [hep-ph]].
%3 citations counted in INSPIRE as of 12 May 2022

%\cite{Giron:2020wpx}
\bibitem{Giron:2020wpx}
J.~F.~Giron and R.~F.~Lebed,
{\it Simple spectrum of $c\bar c c\bar c$ states in the dynamical diquark model},
\href{http://dx.doi.org/10.1103/PhysRevD.102.074003}{Phys. Rev. D \textbf{102} (2020) no.7, 074003}.
%[arXiv:2008.01631 [hep-ph]].
%53 citations counted in INSPIRE as of 12 May 2022

%\cite{Karliner:2020dta}
\bibitem{Karliner:2020dta}
M.~Karliner and J.~L.~Rosner,
{\it Interpretation of structure in the di- $J/\psi$ spectrum},
\href{http://dx.doi.org/10.1103/PhysRevD.102.114039}{Phys. Rev. D \textbf{102} (2020) no.11, 114039}.
%[arXiv:2009.04429 [hep-ph]].
%46 citations counted in INSPIRE as of 12 May 2022

%\cite{Zhao:2020zjh}
\bibitem{Zhao:2020zjh}
Z.~Zhao, K.~Xu, A.~Kaewsnod, X.~Liu, A.~Limphirat and Y.~Yan,
{\it Study of charmoniumlike and fully-charm tetraquark spectroscopy},
\href{http://dx.doi.org/10.1103/PhysRevD.103.116027}{Phys. Rev. D \textbf{103} (2021) no.11, 116027}.
%[arXiv:2012.15554 [hep-ph]].
%19 citations counted in INSPIRE as of 12 May 2022

%\cite{Mutuk:2021hmi}
\bibitem{Mutuk:2021hmi}
H.~Mutuk,
{\it Nonrelativistic treatment of fully-heavy tetraquarks as diquark-antidiquark states},
\href{http://dx.doi.org/10.1140/epjc/s10052-021-09176-8}{Eur. Phys. J. C \textbf{81} (2021) no.4, 367}.
%[arXiv:2104.11823 [hep-ph]].
%8 citations counted in INSPIRE as of 12 May 2022

%\cite{Wang:2021kfv}
\bibitem{Wang:2021kfv}
G.~J.~Wang, L.~Meng, M.~Oka and S.~L.~Zhu,
{\it Higher fully charmed tetraquarks: Radial excitations and P-wave states},
\href{http://dx.doi.org/10.1103/PhysRevD.104.036016}{Phys. Rev. D \textbf{104} (2021) no.3, 036016}.
%[arXiv:2105.13109 [hep-ph]].
%8 citations counted in INSPIRE as of 12 May 2022

%\cite{Wang:2020ols}
\bibitem{Wang:2020ols}
Z.~G.~Wang,
{\it Tetraquark candidates in the LHCb's di-$J/\psi$ mass spectrum},
\href{http://dx.doi.org/10.1088/1674-1137/abb080}{Chin. Phys. C \textbf{44} (2020) no.11, 113106}.
%[arXiv:2006.13028 [hep-ph]].
%38 citations counted in INSPIRE as of 12 May 2022

%\cite{Ke:2021iyh}
\bibitem{Ke:2021iyh}
H.~W.~Ke, X.~Han, X.~H.~Liu and Y.~L.~Shi,
{\it Tetraquark state $X(6900)$ and the interaction between diquark and antidiquark},
\href{http://dx.doi.org/10.1140/epjc/s10052-021-09229-y}{Eur. Phys. J. C \textbf{81} (2021) no.5, 427}.
%[arXiv:2103.13140 [hep-ph]].
%17 citations counted in INSPIRE as of 12 May 2022

%\cite{Zhu:2020xni}
\bibitem{Zhu:2020xni}
R.~Zhu,
{\it Fully-heavy tetraquark spectra and production at hadron colliders},
\href{http://dx.doi.org/10.1016/j.nuclphysb.2021.115393}{Nucl. Phys. B \textbf{966} (2021), 115393}.
%[arXiv:2010.09082 [hep-ph]].
%28 citations counted in INSPIRE as of 12 May 2022

%\cite{Jin:2020jfc}
\bibitem{Jin:2020jfc}
X.~Jin, Y.~Xue, H.~Huang and J.~Ping,
{\it Full-heavy tetraquarks in constituent quark models},
\href{http://dx.doi.org/10.1140/epjc/s10052-020-08650-z}{Eur. Phys. J. C \textbf{80} (2020) no.11, 1083}.
%[arXiv:2006.13745 [hep-ph]].
%48 citations counted in INSPIRE as of 12 May 2022

%\cite{Yang:2021hrb}
\bibitem{Yang:2021hrb}
G.~Yang, J.~Ping and J.~Segovia,
{\it Exotic resonances of fully-heavy tetraquarks in a lattice-QCD insipired quark model},
\href{http://dx.doi.org/10.1103/PhysRevD.104.014006}{Phys. Rev. D \textbf{104} (2021) no.1, 014006}.
%[arXiv:2104.08814 [hep-ph]].
%11 citations counted in INSPIRE as of 12 May 2022

%\cite{Albuquerque:2020hio}
\bibitem{Albuquerque:2020hio}
R.~M.~Albuquerque, S.~Narison, A.~Rabemananjara, D.~Rabetiarivony and G.~Randriamanatrika,
{\it Doubly-hidden scalar heavy molecules and tetraquarks states from QCD at NLO},
\href{http://dx.doi.org/10.1103/PhysRevD.102.094001}{Phys. Rev. D \textbf{102} (2020) no.9, 094001}.
%[arXiv:2008.01569 [hep-ph]].
%40 citations counted in INSPIRE as of 12 May 2022

%\cite{Albuquerque:2021erv}
\bibitem{Albuquerque:2021erv}
R.~M.~Albuquerque, S.~Narison, D.~Rabetiarivony and G.~Randriamanatrika,
{\it Doubly hidden $0^{++}$ molecules and tetraquarks states from QCD at NLO},
\href{http://dx.doi.org/10.1016/j.nuclphysbps.2021.05.031}{Nucl. Part. Phys. Proc. \textbf{312-317} (2021), 120-124}.
%[arXiv:2102.08776 [hep-ph]].
%5 citations counted in INSPIRE as of 12 May 2022

%\cite{Wu:2022qwd}
\bibitem{Wu:2022qwd}
R.~H.~Wu, Y.~S.~Zuo, C.~Y.~Wang, C.~Meng, Y.~Q.~Ma and K.~T.~Chao,
{\it NLO results with operator mixing for fully heavy tetraquarks in QCD sum rules},
\href{http://arxiv.org/abs/arXiv:2201.11714}{arXiv:2201.11714 [hep-ph]}.
%2 citations counted in INSPIRE as of 12 May 2022

%\cite{Asadi:2021ids}
\bibitem{Asadi:2021ids}
Z.~Asadi and G.~R.~Boroun,
{\it Masses of fully heavy tetraquark states from a four-quark static potential model},
\href{http://dx.doi.org/10.1103/PhysRevD.105.014006}{Phys. Rev. D \textbf{105} (2022) no.1, 014006}.
%[arXiv:2112.11028 [hep-ph]].
%1 citations counted in INSPIRE as of 12 May 2022

%\cite{Yang:2020wkh}
\bibitem{Yang:2020wkh}
B.~C.~Yang, L.~Tang and C.~F.~Qiao,
{\it Scalar fully-heavy tetraquark states $QQ^\prime {\bar{Q}} \bar{Q^\prime }$ in QCD sum rules},
\href{http://dx.doi.org/10.1140/epjc/s10052-021-09096-7}{Eur. Phys. J. C \textbf{81} (2021) no.4, 324}.
%[arXiv:2012.04463 [hep-ph]].
%18 citations counted in INSPIRE as of 12 May 2022

%\cite{Huang:2020dci}
\bibitem{Huang:2020dci}
G.~Huang, J.~Zhao and P.~Zhuang,
{\it Pair structure of heavy tetraquark systems},
\href{http://dx.doi.org/10.1103/PhysRevD.103.054014}{Phys. Rev. D \textbf{103} (2021) no.5, 054014}.
%[arXiv:2012.14845 [hep-ph]].
%7 citations counted in INSPIRE as of 12 May 2022

%\cite{Feng:2020riv}
\bibitem{Feng:2020riv}
F.~Feng, Y.~Huang, Y.~Jia, W.~L.~Sang, X.~Xiong and J.~Y.~Zhang,
{\it Fragmentation production of fully-charmed tetraquarks at LHC},
\href{http://arxiv.org/abs/arXiv:2009.08450}{arXiv:2009.08450 [hep-ph]}.
%21 citations counted in INSPIRE as of 12 May 2022

%\cite{Huang:2021vtb}
\bibitem{Huang:2021vtb}
Y.~Huang, F.~Feng, Y.~Jia, W.~L.~Sang, D.~S.~Yang and J.~Y.~Zhang,
{\it Inclusive production of fully-charmed $1^{+-}$ tetraquark at B factory},
\href{http://dx.doi.org/10.1088/1674-1137/ac0b38}{Chin. Phys. C \textbf{45} (2021) no.9, 093101}.
%[arXiv:2104.03887 [hep-ph]].
%6 citations counted in INSPIRE as of 12 May 2022

%\cite{Ma:2020kwb}
\bibitem{Ma:2020kwb}
Y.~Q.~Ma and H.~F.~Zhang,
{\it Exploring the Di-$J/\psi$ Resonances around 6.9 $\mathrm{GeV}$ Based on $ab$ $initio$ Perturbative QCD},
\href{http://arxiv.org/abs/arXiv:2009.08376}{arXiv:2009.08376 [hep-ph]}.
%26 citations counted in INSPIRE as of 12 May 2022

%\cite{Maciula:2020wri}
\bibitem{Maciula:2020wri}
R.~Maciu\l{}a, W.~Sch\"afer and A.~Szczurek,
{\it On the mechanism of $T_{4c}$(6900) tetraquark production},
\href{http://dx.doi.org/10.1016/j.physletb.2020.136010}{Phys. Lett. B \textbf{812} (2021), 136010}.
%[arXiv:2009.02100 [hep-ph]].
%33 citations counted in INSPIRE as of 12 May 2022

%\cite{Goncalves:2021ytq}
\bibitem{Goncalves:2021ytq}
V.~P.~Gon\c{c}alves and B.~D.~Moreira,
{\it Fully - heavy tetraquark production by $\gamma\gamma$ interactions in hadronic collisions at the LHC},
\href{http://dx.doi.org/10.1016/j.physletb.2021.136249}{Phys. Lett. B \textbf{816} (2021), 136249}.
%[arXiv:2101.03798 [hep-ph]].
%10 citations counted in INSPIRE as of 12 May 2022

%\cite{Wang:2020gmd}
\bibitem{Wang:2020gmd}
X.~Y.~Wang, Q.~Y.~Lin, H.~Xu, Y.~P.~Xie, Y.~Huang and X.~Chen,
{\it Discovery potential for the LHCb fully-charm tetraquark $X(6900)$ state via $\bar{p}p$ annihilation reaction},
\href{http://dx.doi.org/10.1103/PhysRevD.102.116014}{Phys. Rev. D \textbf{102} (2020), 116014}.
%[arXiv:2007.09697 [hep-ph]].
%27 citations counted in INSPIRE as of 12 May 2022

%\cite{Esposito:2021ptx}
\bibitem{Esposito:2021ptx}
A.~Esposito, C.~A.~Manzari, A.~Pilloni and A.~D.~Polosa,
{\it Hunting for tetraquarks in ultraperipheral heavy ion collisions},
\href{http://dx.doi.org/10.1103/PhysRevD.104.114029}{Phys. Rev. D \textbf{104} (2021) no.11, 114029}.
%[arXiv:2109.10359 [hep-ph]].
%6 citations counted in INSPIRE as of 12 May 2022

%\cite{Zhuang:2021pci}
\bibitem{Zhuang:2021pci}
Z.~Zhuang, Y.~Zhang, Y.~Ma and Q.~Wang,
{\it Lineshape of the compact fully heavy tetraquark},
\href{http://dx.doi.org/10.1103/PhysRevD.105.054026}{Phys. Rev. D \textbf{105} (2022) no.5, 054026}.
%[arXiv:2111.14028 [hep-ph]].
%1 citations counted in INSPIRE as of 12 May 2022

%\cite{Zhao:2020nwy}
\bibitem{Zhao:2020nwy}
J.~Zhao, S.~Shi and P.~Zhuang,
{\it Fully-heavy tetraquarks in a strongly interacting medium},
\href{http://dx.doi.org/10.1103/PhysRevD.102.114001}{Phys. Rev. D \textbf{102} (2020) no.11, 114001}.
%[arXiv:2009.10319 [hep-ph]].
%28 citations counted in INSPIRE as of 12 May 2022

%\cite{Becchi:2020uvq}
\bibitem{Becchi:2020uvq}
C.~Becchi, J.~Ferretti, A.~Giachino, L.~Maiani and E.~Santopinto,
{\it A study of $c c\bar{c}\bar{c}$ tetraquark decays in 4 muons and in $D^{(*)} \bar{D}^{(*)}$ at LHC},
\href{http://dx.doi.org/10.1016/j.physletb.2020.135952}{Phys. Lett. B \textbf{811} (2020), 135952}.
%[arXiv:2006.14388 [hep-ph]].
%37 citations counted in INSPIRE as of 12 May 2022

%\cite{Andrade:2022rbn}
\bibitem{Andrade:2022rbn}
S.~Andrad\'e, M.~Siddikov and I.~Schmidt,
{\it Exclusive photoproduction of heavy quarkonia pairs},
\href{http://dx.doi.org/10.1103/PhysRevD.105.076022}{Phys. Rev. D \textbf{105} (2022) no.7, 076022}.
%[arXiv:2202.03288 [hep-ph]].
%1 citations counted in INSPIRE as of 12 May 2022

%\cite{Lansberg:2020ejc}
\bibitem{Lansberg:2020ejc}
J.~P.~Lansberg and M.~A.~Ozcelik,
{\it Curing the unphysical behaviour of NLO quarkonium production at the LHC and its relevance to constrain the gluon PDF at low scales},
\href{http://dx.doi.org/10.1140/epjc/s10052-021-09258-7}{Eur. Phys. J. C \textbf{81} (2021) no.6, 497}.
%[arXiv:2012.00702 [hep-ph]].
%10 citations counted in INSPIRE as of 12 May 2022

%\cite{Sonnenschein:2020nwn}
\bibitem{Sonnenschein:2020nwn}
J.~Sonnenschein and D.~Weissman,
{\it Deciphering the recently discovered tetraquark candidates around 6.9 GeV},
\href{http://dx.doi.org/10.1140/epjc/s10052-020-08818-7}{Eur. Phys. J. C \textbf{81} (2021) no.1, 25}.
%[arXiv:2008.01095 [hep-ph]].
%31 citations counted in INSPIRE as of 12 May 2022

%\cite{Zhu:2020snb}
\bibitem{Zhu:2020snb}
J.~W.~Zhu, X.~D.~Guo, R.~Y.~Zhang, W.~G.~Ma and X.~Q.~Li,
{\it A possible interpretation for $X(6900)$ observed in four-muon final state by LHCb -- A light Higgs-like boson?}
\href{http://arxiv.org/abs/arXiv:2011.07799}{arXiv:2011.07799 [hep-ph]}.
%17 citations counted in INSPIRE as of 12 May 2022

%\cite{Wan:2020fsk}
\bibitem{Wan:2020fsk}
B.~D.~Wan and C.~F.~Qiao,
{\it Gluonic tetracharm configuration of $X (6900)$},
\href{http://dx.doi.org/10.1016/j.physletb.2021.136339}{Phys. Lett. B \textbf{817} (2021), 136339}.
%[arXiv:2012.00454 [hep-ph]].
%21 citations counted in INSPIRE as of 12 May 2022

%\cite{Gordillo:2020sgc}
\bibitem{Gordillo:2020sgc}
M.~C.~Gordillo, F.~De Soto and J.~Segovia,
{\it Diffusion Monte Carlo calculations of fully-heavy multiquark bound states},
\href{http://dx.doi.org/10.1103/PhysRevD.102.114007}{Phys. Rev. D \textbf{102} (2020) no.11, 114007}.
%[arXiv:2009.11889 [hep-ph]].
%26 citations counted in INSPIRE as of 12 May 2022

%\cite{Liu:2020tqy}
\bibitem{Liu:2020tqy}
M.~Z.~Liu and L.~S.~Geng,
{\it Is $X(7200)$ the heavy anti-quark diquark symmetry partner of $ X(3872)$?}
\href{http://dx.doi.org/10.1140/epjc/s10052-021-08980-6}{Eur. Phys. J. C \textbf{81} (2021) no.2, 179}.
%[arXiv:2012.05096 [hep-ph]].
%8 citations counted in INSPIRE as of 12 May 2022

%\cite{Majarshin:2021hex}
\bibitem{Majarshin:2021hex}
A.~J.~Majarshin, Y.~A.~Luo, F.~Pan and J.~Segovia,
{\it Bosonic algebraic approach applied to the [QQ][$\bar{Q}\bar{Q}$] tetraquarks},
\href{http://dx.doi.org/10.1103/PhysRevD.105.054024}{Phys. Rev. D \textbf{105} (2022) no.5, 054024}.
%[arXiv:2106.01179 [hep-ph]].
%3 citations counted in INSPIRE as of 12 May 2022

%\cite{Sombillo:2021rxv}
\bibitem{Sombillo:2021rxv}
D.~L.~B.~Sombillo, Y.~Ikeda, T.~Sato and A.~Hosaka,
{\it Model independent analysis of coupled-channel scattering: A deep learning approach},
\href{http://dx.doi.org/10.1103/PhysRevD.104.036001}{Phys. Rev. D \textbf{104} (2021) no.3, 036001}.
%[arXiv:2105.04898 [hep-ph]].
%10 citations counted in INSPIRE as of 12 May 2022

%\cite{Kuang:2022vdy}
\bibitem{Kuang:2022vdy}
Z.~Kuang, K.~Serafin, X.~Zhao and J.~P.~Vary,
{\it All-charm tetraquark in Front Form dynamics},
\href{http://dx.doi.org/10.1103/PhysRevD.105.094028}{Phys. Rev. D \textbf{105} (2022), 094028}.

%\cite{Yang:2020rih}
\bibitem{Yang:2020rih}
G.~Yang, J.~Ping, L.~He and Q.~Wang,
{\it Potential model prediction of fully-heavy tetraquarks $QQ\bar{Q}\bar{Q}$ ($Q=c, b$)},
\href{http://arxiv.org/abs/arXiv:2006.13756}{arXiv:2006.13756 [hep-ph]}.
%39 citations counted in INSPIRE as of 13 May 2022

%\cite{Yang:2021zrc}
\bibitem{Yang:2021zrc}
Z.~H.~Yang, Q.~N.~Wang, W.~Chen and H.~X.~Chen,
{\it Investigation of the stability for fully-heavy $bc\bar{b}\bar{c}$ tetraquark states},
\href{http://dx.doi.org/10.1103/PhysRevD.104.014003}{Phys. Rev. D \textbf{104} (2021) no.1, 014003}.
%[arXiv:2102.10605 [hep-ph]].
%4 citations counted in INSPIRE as of 13 May 2022

%\cite{Wang:2021taf}
\bibitem{Wang:2021taf}
Q.~N.~Wang, Z.~Y.~Yang, W.~Chen and H.~X.~Chen,
{\it Mass spectra for the $cc\bar b \bar b$ and $bb \bar c\bar c$ tetraquark states},
\href{http://dx.doi.org/10.1103/PhysRevD.104.014020}{Phys. Rev. D \textbf{104} (2021) no.1, 014020}.
%[arXiv:2106.05550 [hep-ph]].
%2 citations counted in INSPIRE as of 13 May 2022

%\cite{Wang:2021mma}
\bibitem{Wang:2021mma}
Q.~N.~Wang, Z.~Y.~Yang and W.~Chen,
{\it Exotic fully-heavy $Q\bar QQ\bar Q$ tetraquark states in $\mathbf{8}_{[Q\bar{Q}]}\otimes \mathbf{8}_{[Q\bar{Q}]}$ color configuration},
\href{http://dx.doi.org/10.1103/PhysRevD.104.114037}{Phys. Rev. D \textbf{104} (2021) no.11, 114037}.
%[arXiv:2109.08091 [hep-ph]].
%2 citations counted in INSPIRE as of 13 May 2022

%\cite{Chen:2022asf}
\bibitem{Chen:2022asf}
H.~X.~Chen, W.~Chen, X.~Liu, Y.~R.~Liu and S.~L.~Zhu,
{\it An updated review of the new hadron states},
\href{http://arxiv.org/abs/arXiv:2204.02649}{arXiv:2204.02649 [hep-ph]}.
%0 citations counted in INSPIRE as of 20 Apr 2022

%\cite{Chen:2016qju}
\bibitem{Chen:2016qju}
H.~X.~Chen, W.~Chen, X.~Liu and S.~L.~Zhu,
{\it The hidden-charm pentaquark and tetraquark states},
\href{http://dx.doi.org/10.1016/j.physrep.2016.05.004}{Phys. Rept. \textbf{639}, 1-121 (2016)}.

%\cite{Liu:2019zoy}
\bibitem{Liu:2019zoy}
  Y.~R.~Liu, H.~X.~Chen, W.~Chen, X.~Liu and S.~L.~Zhu,
  {\it Pentaquark and Tetraquark States},
  \href{http://dx.doi.org/10.1016/j.ppnp.2019.04.003}{Prog.\ Part.\ Nucl.\ Phys.\  {\bf 107}, 237 (2019)}.
  %%CITATION = doi:10.1016/j.ppnp.2019.04.003;%%
  %42 citations counted in INSPIRE as of 22 Sep 2019

%\cite{Lebed:2016hpi}
\bibitem{Lebed:2016hpi}
R.~F.~Lebed, R.~E.~Mitchell and E.~S.~Swanson,
{\it Heavy-quark QCD exotica},
\href{http://dx.doi.org/10.1016/j.ppnp.2016.11.003}{Prog. Part. Nucl. Phys. \textbf{93}, 143-194 (2017)}.
%[arXiv:1610.04528 [hep-ph]].
%415 citations counted in INSPIRE as of 22 Apr 2022

%\cite{Esposito:2016noz}
\bibitem{Esposito:2016noz}
A.~Esposito, A.~Pilloni and A.~D.~Polosa,
{\it Multiquark resonances},
\href{http://dx.doi.org/10.1016/j.physrep.2016.11.002}{Phys. Rept. \textbf{668}, 1-97 (2017)}.
%[arXiv:1611.07920 [hep-ph]].
%459 citations counted in INSPIRE as of 20 Apr 2022

%\cite{Hosaka:2016pey}
\bibitem{Hosaka:2016pey}
A.~Hosaka, T.~Iijima, K.~Miyabayashi, Y.~Sakai and S.~Yasui,
{\it Exotic hadrons with heavy flavors: X, Y, Z, and related states},
\href{http://dx.doi.org/10.1093/ptep/ptw045}{PTEP \textbf{2016} no.6, 062C01 (2016)}.
%[arXiv:1603.09229 [hep-ph]].
%187 citations counted in INSPIRE as of 20 Apr 2022

%\cite{Guo:2017jvc}
\bibitem{Guo:2017jvc}
F.~K.~Guo, C.~Hanhart, U.~G.~Meissner, Q.~Wang, Q.~Zhao and B.~S.~Zou,
{\it Hadronic molecules},
\href{http://dx.doi.org/10.1103/RevModPhys.90.015004}{Rev. Mod. Phys. \textbf{90} no.1, 015004 (2018)}.
%[arXiv:1705.00141 [hep-ph]].
%720 citations counted in INSPIRE as of 20 Apr 2022

%\cite{Ali:2017jda}
\bibitem{Ali:2017jda}
A.~Ali, J.~S.~Lange and S.~Stone,
{\it Exotics: Heavy pentaquarks and tetraquarks},
\href{http://dx.doi.org/10.1016/j.ppnp.2017.08.003}{Prog. Part. Nucl. Phys. \textbf{97} , 123-198 (2017)}.
%[arXiv:1706.00610 [hep-ph]].
%347 citations counted in INSPIRE as of 20 Apr 2022

%\cite{Olsen:2017bmm}
\bibitem{Olsen:2017bmm}
S.~L.~Olsen, T.~Skwarnicki and D.~Zieminska,
{\it Nonstandard heavy mesons and baryons: Experimental evidence},
\href{http://dx.doi.org/10.1103/RevModPhys.90.015003}{Rev. Mod. Phys. \textbf{90} no.1, 015003 (2018)}.
%[arXiv:1708.04012 [hep-ph]]..
%447 citations counted in INSPIRE as of 20 Apr 2022

%\cite{Karliner:2017qhf}
\bibitem{Karliner:2017qhf}
M.~Karliner, J.~L.~Rosner and T.~Skwarnicki,
{\it Multiquark States},
\href{http://dx.doi.org/10.1146/annurev-nucl-101917-020902}{Ann. Rev. Nucl. Part. Sci. \textbf{68}, 17-44 (2018)}.
%[arXiv:1711.10626 [hep-ph]].
%152 citations counted in INSPIRE as of 20 Apr 2022

%\cite{Bass:2018xmz}
\bibitem{Bass:2018xmz}
S.~D.~Bass and P.~Moskal,
{\it $\eta^\prime$ and $\eta$ mesons with connection to anomalous glue},
\href{http://dx.doi.org/10.1103/RevModPhys.91.015003}{Rev. Mod. Phys. \textbf{91} no.1, 015003 (2019)}.
%[arXiv:1810.12290 [hep-ph]].
%39 citations counted in INSPIRE as of 20 Apr 2022

%\cite{Brambilla:2019esw}
\bibitem{Brambilla:2019esw}
N.~Brambilla, S.~Eidelman, C.~Hanhart, A.~Nefediev, C.~P.~Shen, C.~E.~Thomas, A.~Vairo and C.~Z.~Yuan,
{\it The $XYZ$ states: Experimental and theoretical status and perspectives},
\href{http://dx.doi.org/10.1016/j.physrep.2020.05.001}{Phys. Rept. \textbf{873}, 1-154 (2020)}.
%[arXiv:1907.07583 [hep-ex]].
%332 citations counted in INSPIRE as of 20 Apr 2022

%\cite{Guo:2019twa}
\bibitem{Guo:2019twa}
F.~K.~Guo, X.~H.~Liu and S.~Sakai,
{\it Threshold cusps and triangle singularities in hadronic reactions},
\href{http://dx.doi.org/10.1016/j.ppnp.2020.103757}{Prog. Part. Nucl. Phys. \textbf{112}, 103757 (2020)}.
%[arXiv:1912.07030 [hep-ph]].
%118 citations counted in INSPIRE as of 20 Apr 2022

%\cite{Ketzer:2019wmd}
\bibitem{Ketzer:2019wmd}
B.~Ketzer, B.~Grube and D.~Ryabchikov,
{\it Light-meson spectroscopy with COMPASS},
\href{http://dx.doi.org/10.1016/j.ppnp.2020.103755}{Prog. Part. Nucl. Phys. \textbf{113}, 103755 (2020)}.
%[arXiv:1909.06366 [hep-ex]].
%21 citations counted in INSPIRE as of 20 Apr 2022

%\cite{Yang:2020atz}
\bibitem{Yang:2020atz}
G.~Yang, J.~Ping and J.~Segovia,
{\it Tetra- and Penta-quark Structures in the Constituent Quark Model},
\href{http://dx.doi.org/10.3390/sym12111869}{Symmetry \textbf{12} no.11, 1869 (2020)}.
%[arXiv:2009.00238 [hep-ph]].
%38 citations counted in INSPIRE as of 20 Apr 2022

%\cite{Roberts:2021nhw}
\bibitem{Roberts:2021nhw}
C.~D.~Roberts, D.~G.~Richards, T.~Horn and L.~Chang,
{\it Insights into the emergence of mass from studies of pion and kaon structure},
\href{http://dx.doi.org/10.1016/j.ppnp.2021.103883}{Prog. Part. Nucl. Phys. \textbf{120}, 103883 (2021)}.
%[arXiv:2102.01765 [hep-ph]].
%47 citations counted in INSPIRE as of 20 Apr 2022

%\cite{Fang:2021wes}
\bibitem{Fang:2021wes}
S.~S.~Fang, B.~Kubis and A.~Kupsc,
{\it What can we learn about light-meson interactions at electron-positron colliders?}
\href{http://dx.doi.org/10.1016/j.ppnp.2021.103884}{Prog. Part. Nucl. Phys. \textbf{120}, 103884 (2021)}.
%[arXiv:2102.05922 [hep-ph]].
%5 citations counted in INSPIRE as of 20 Apr 2022

%\cite{Jin:2021vct}
\bibitem{Jin:2021vct}
S.~Jin and X.~Shen,
{\it Highlights of light meson spectroscopy at the BESIII experiment},
\href{http://dx.doi.org/10.1093/nsr/nwab198}{Natl. Sci. Rev. \textbf{8} no.11, nwab198 (2021)}.
%1 citations counted in INSPIRE as of 20 Apr 2022

%\cite{JPAC:2021rxu}
\bibitem{JPAC:2021rxu}
M.~Albaladejo \textit{et al.} [JPAC Collaboration],
{\it Novel approaches in Hadron Spectroscopy},
\href{http://arxiv.org/abs/arXiv:2112.13436}{arXiv:2112.13436 [hep-ph]}.
%5 citations counted in INSPIRE as of 20 Apr 2022

%\cite{Meng:2022ozq}
\bibitem{Meng:2022ozq}
L.~Meng, B.~Wang, G.~J.~Wang and S.~L.~Zhu,
{\it Chiral perturbation theory for heavy hadrons and chiral effective field theory for heavy hadronic molecules},
\href{http://arxiv.org/abs/arXiv:2204.08716}{arXiv:2204.08716 [hep-ph]}.
%0 citations counted in INSPIRE as of 20 Apr 2022

%\cite{Mai:2022eur}
\bibitem{Mai:2022eur}
M.~Mai, U.~G.~Mei\ss{}ner and C.~Urbach,
{\it Towards a theory of hadron resonances},
\href{http://arxiv.org/abs/arXiv:2206.01477}{[arXiv:2206.01477 [hep-ph]]}.
%1 citations counted in INSPIRE as of 16 Jul 2022

%\cite{Maiani:2022psl}
\bibitem{Maiani:2022psl}
L.~Maiani and A.~Pilloni,
{\it GGI Lectures on Exotic Hadrons},
\href{http://arxiv.org/abs/arXiv:2207.05141}{[arXiv:2207.05141 [hep-ph]]}.
%0 citations counted in INSPIRE as of 16 Jul 2022

%\cite{Maiani:2020pur}
\bibitem{Maiani:2020pur}
L.~Maiani,
{\it $J/\psi$-pair resonance by LHCb: a new revolution?}
\href{http://dx.doi.org/10.1016/j.scib.2020.08.019}{Sci. Bull. \textbf{65} (2020), 1949-1951}.
%[arXiv:2008.01637 [hep-ph]].
%23 citations counted in INSPIRE as of 12 May 2022

%\cite{Chao:2020dml}
\bibitem{Chao:2020dml}
K.~T.~Chao and S.~L.~Zhu,
{\it The possible tetraquark states $cc \bar c \bar c$ observed by the LHCb experiment},
\href{http://dx.doi.org/10.1016/j.scib.2020.08.031}{Sci. Bull. \textbf{65} (2020) no.23, 1952-1953}.
%[arXiv:2008.07670 [hep-ph]].
%30 citations counted in INSPIRE as of 12 May 2022

%\cite{Richard:2020hdw}
\bibitem{Richard:2020hdw}
J.~M.~Richard,
{\it About the $J/\psi$ $J/\psi$ peak of LHCb: fully-charmed tetraquark?}
\href{http://dx.doi.org/10.1016/j.scib.2020.08.020}{Sci. Bull. \textbf{65} (2020), 1954-1955}.
%[arXiv:2008.01962 [hep-ph]].
%29 citations counted in INSPIRE as of 12 May 2022

%\cite{Wang:2020wrp}
\bibitem{Wang:2020wrp}
J.~Z.~Wang, D.~Y.~Chen, X.~Liu and T.~Matsuki,
{\it Producing fully charm structures in the $J/\psi$ -pair invariant mass spectrum},
\href{http://dx.doi.org/10.1103/PhysRevD.103.L071503}{Phys. Rev. D \textbf{103} (2021) no.7, 071503}.
%[arXiv:2008.07430 [hep-ph]].
%34 citations counted in INSPIRE as of 12 May 2022

%\cite{Dong:2020nwy}
\bibitem{Dong:2020nwy}
X.~K.~Dong, V.~Baru, F.~K.~Guo, C.~Hanhart and A.~Nefediev,
{\it Coupled-Channel Interpretation of the LHCb Double-$J/\psi$ Spectrum and Hints of a New State Near the $J/\psi J/\psi$ Threshold},
\href{http://dx.doi.org/10.1103/PhysRevLett.126.132001}{Phys. Rev. Lett. \textbf{126} (2021) no.13, 132001}
[erratum: Phys. Rev. Lett. \textbf{127} (2021) no.11, 119901].

%\cite{Wang:2020tpt}
\bibitem{Wang:2020tpt}
J.~Z.~Wang, X.~Liu and T.~Matsuki,
{\it Fully-heavy structures in the invariant mass spectrum of $J/\psi \psi(3686)$, $J/\psi \psi(3770)$, $\psi(3686) \psi(3686)$, and $J/\psi \Upsilon(1S)$ at hadron colliders},
\href{http://dx.doi.org/10.1016/j.physletb.2021.136209}{Phys. Lett. B \textbf{816} (2021), 136209}.
%[arXiv:2012.03281 [hep-ph]].
%8 citations counted in INSPIRE as of 12 May 2022

%\cite{Dong:2020hxe}
\bibitem{Dong:2020hxe}
X.~K.~Dong, F.~K.~Guo and B.~S.~Zou,
{\it Explaining the Many Threshold Structures in the Heavy-Quark Hadron Spectrum},
\href{http://dx.doi.org/10.1103/PhysRevLett.126.152001}{Phys. Rev. Lett. \textbf{126} (2021) no.15, 152001}.


%\cite{Liang:2021fzr}
\bibitem{Liang:2021fzr}
Z.~R.~Liang, X.~Y.~Wu and D.~L.~Yao,
{\it Hunting for states in the recent LHCb di-J/\ensuremath{\psi} invariant mass spectrum},
\href{http://dx.doi.org/10.1103/PhysRevD.104.034034}{Phys. Rev. D \textbf{104} (2021) no.3, 034034}.
%[arXiv:2104.08589 [hep-ph]].
%11 citations counted in INSPIRE as of 12 May 2022

%\cite{Nefediev:2021pww}
\bibitem{Nefediev:2021pww}
A.~V.~Nefediev,
{\it X(6200) as a compact tetraquark in the QCD string model},
\href{http://dx.doi.org/10.1140/epjc/s10052-021-09511-z}{Eur. Phys. J. C \textbf{81} (2021) no.8, 692}.
%[arXiv:2107.14182 [hep-ph]].
%3 citations counted in INSPIRE as of 12 May 2022

%\cite{Gong:2020bmg}
\bibitem{Gong:2020bmg}
C.~Gong, M.~C.~Du, Q.~Zhao, X.~H.~Zhong and B.~Zhou,
{\it Nature of X(6900) and its production mechanism at LHCb},
\href{http://dx.doi.org/10.1016/j.physletb.2021.136794}{Phys. Lett. B \textbf{824} (2022), 136794}.
%[arXiv:2011.11374 [hep-ph]].
%20 citations counted in INSPIRE as of 12 May 2022

%\cite{Dong:2021lkh}
\bibitem{Dong:2021lkh}
X.~K.~Dong, V.~Baru, F.~K.~Guo, C.~Hanhart, A.~Nefediev and B.~S.~Zou,
{\it Is the existence of a J/\ensuremath{\psi}J/\ensuremath{\psi} bound state plausible?}
\href{http://dx.doi.org/10.1016/j.scib.2021.09.009}{Sci. Bull. \textbf{66} (2021) no.24, 2462-2470}.
%[arXiv:2107.03946 [hep-ph]].
%8 citations counted in INSPIRE as of 12 May 2022

%\cite{Guo:2020pvt}
\bibitem{Guo:2020pvt}
Z.~H.~Guo and J.~A.~Oller,
{\it Insights into the inner structures of the fully charmed tetraquark state $X(6900)$},
\href{http://dx.doi.org/10.1103/PhysRevD.103.034024}{Phys. Rev. D \textbf{103} (2021) no.3, 034024}.
%[arXiv:2011.00978 [hep-ph]].
%27 citations counted in INSPIRE as of 12 May 2022

%\cite{Cao:2020gul}
\bibitem{Cao:2020gul}
Q.~F.~Cao, H.~Chen, H.~R.~Qi and H.~Q.~Zheng,
{\it Some remarks on X(6900)},
\href{http://dx.doi.org/10.1088/1674-1137/ac0ee5}{Chin. Phys. C \textbf{45} (2021) no.10, 103102}.
%[arXiv:2011.04347 [hep-ph]].
%22 citations counted in INSPIRE as of 12 May 2022

%\cite{Wang:2022jmb}
\bibitem{Wang:2022jmb}
J.~Z.~Wang and X.~Liu,
{\it Improved understanding of the peaking phenomenon existing in the di-$J/\psi$ invariant mass spectrum newly from the CMS Collaboration},
\href{http://dx.doi.org/10.1103/PhysRevD.106.054015}{Phys. Rev. D \textbf{106} (2022) no.5, 054015}.

%\cite{Zhou:2022xpd}
\bibitem{Zhou:2022xpd}
Q.~Zhou, D.~Guo, S.~Q.~Kuang, Q.~H.~Yang and L.~Y.~Dai,
{\it Nature of the $X(6900)$ in partial wave decomposition of $J/\psi J/\psi$ scattering},
\href{http://arxiv.org/abs/arXiv:2207.07537}{[arXiv:2207.07537 [hep-ph]]}.
%0 citations counted in INSPIRE as of 18 Jul 2022

\bibitem{CMS}
Kai Yi on behalf of the CMS Collaboration, Recent CMS results on exotic resonance, Proceedings at ICHEP 2022, \href{https://agenda.infn.it/event/28874/contributions/170300/}{https://agenda.infn.it/event/28874/contributions/170300/}.

\bibitem{ATLAS}
Evelina Bouhova-Thacker on behalf of the ATLAS Collaboration, ATLAS results on exotic hadronic resonances, Proceedings at ICHEP 2022,
\href{https://agenda.infn.it/event/28874/contributions/170298/}{https://agenda.infn.it/event/28874/contributions/170298/}.

%\cite{Chao:1980dv}
\bibitem{Chao:1980dv}
  K.~T.~Chao,
  {\it The $(cc)$-$(\bar c \bar c)$ (Diquark-Antidiquark) States in $e^+ e^-$ Annihilation},
\href{http://dx.doi.org/10.1007/BF01431564}{Z.\ Phys.\ C {\bf 7}, 317 (1981)}.
  %%CITATION = doi:10.1007/BF01431564;%%
  %26 citations counted in INSPIRE as of 26 Jun 2020

%\cite{Iwasaki:1975pv}
\bibitem{Iwasaki:1975pv}
  Y.~Iwasaki,
  {\it A Possible Model for New Resonances-Exotics and Hidden Charm},
\href{http://dx.doi.org/10.1143/PTP.54.492}{Prog.\ Theor.\ Phys.\  {\bf 54}, 492 (1975)}.
  %%CITATION = doi:10.1143/PTP.54.492;%%
  %39 citations counted in INSPIRE as of 26 Jun 2020

%\cite{Ader:1981db}
\bibitem{Ader:1981db}
  J.~P.~Ader, J.~M.~Richard and P.~Taxil,
  {\it Do narrow heavy multi-Quark states exist?}
\href{http://dx.doi.org/10.1103/PhysRevD.25.2370}{Phys.\ Rev.\ D {\bf 25}, 2370 (1982)}.
  %%CITATION = doi:10.1103/PhysRevD.25.2370;%%
  %178 citations counted in INSPIRE as of 26 Jun 2020

%\cite{Heller:1985cb}
\bibitem{Heller:1985cb}
  L.~Heller and J.~A.~Tjon,
  {\it  Bound states of heavy $Q^2 \bar{Q}^2$ systems},
\href{http://dx.doi.org/10.1103/PhysRevD.32.755}{Phys.\ Rev.\ D {\bf 32}, 755 (1985)}.
  %%CITATION = doi:10.1103/PhysRevD.32.755;%%
  %49 citations counted in INSPIRE as of 26 Jun 2020

%\cite{Badalian:1985es}
\bibitem{Badalian:1985es}
  A.~M.~Badalian, B.~L.~Ioffe and A.~V.~Smilga,
  {\it Four-quark states in heavy quark systems},
\href{http://dx.doi.org/10.1016/0550-3213(87)90248-3}{Nucl.\ Phys.\ B {\bf 281}, 85 (1987)}.
  %%CITATION = doi:10.1016/0550-3213(87)90248-3;%%
  %33 citations counted in INSPIRE as of 26 Jun 2020

%\cite{Zouzou:1986qh}
\bibitem{Zouzou:1986qh}
  S.~Zouzou, B.~Silvestre-Brac, C.~Gignoux and J.~M.~Richard,
  {\it Four-quark bound states},
\href{http://dx.doi.org/10.1007/BF01557611}{Z.\ Phys.\ C {\bf 30}, 457 (1986)}.
  %%CITATION = doi:10.1007/BF01557611;%%
  %167 citations counted in INSPIRE as of 26 Jun 2020

%\cite{Lloyd:2003yc}
\bibitem{Lloyd:2003yc}
  R.~J.~Lloyd and J.~P.~Vary,
  {\it All-charm tetraquarks},
\href{http://dx.doi.org/10.1103/PhysRevD.70.014009}{Phys.\ Rev.\ D {\bf 70}, 014009 (2004)}.
  %%CITATION = doi:10.1103/PhysRevD.70.014009;%%
  %39 citations counted in INSPIRE as of 26 Jun 2020

\bibitem{Barnea:2006sd}
  N.~Barnea, J.~Vijande and A.~Valcarce,
  {\it Four-quark spectroscopy within the hyperspherical formalism},
\href{http://dx.doi.org/10.1103/PhysRevD.73.054004}{Phys.\ Rev.\ D {\bf 73}, 054004 (2006)}.
  %%CITATION = doi:10.1103/PhysRevD.73.054004;%%
  %73 citations counted in INSPIRE as of 26 Jun 2020

%\cite{Berezhnoy:2011xn}
\bibitem{Berezhnoy:2011xn}
  A.~V.~Berezhnoy, A.~V.~Luchinsky and A.~A.~Novoselov,
  {\it Heavy tetraquarks production at the LHC},
\href{http://dx.doi.org/10.1103/PhysRevD.86.034004}{Phys.\ Rev.\ D {\bf 86}, 034004 (2012)}.
  %%CITATION = doi:10.1103/PhysRevD.86.034004;%%
  %48 citations counted in INSPIRE as of 27 Jul 2020

%\cite{Yi:2018fxo}
\bibitem{Yi:2018fxo}
  K.~Yi,
  {\it Things that go bump in the night: From $J/\psi\phi$ to other mass spectrum},
\href{http://dx.doi.org/10.1142/S0217751X1850224X}{Int.\ J.\ Mod.\ Phys.\ A {\bf 33}, 1850224 (2019)}.
  %%CITATION = doi:10.1142/S0217751X1850224X;%%
  %4 citations counted in INSPIRE as of 26 Jun 2020

%\cite{Durgut:2018lmn}
\bibitem{Durgut:2018lmn}
S.~Durgut,
{\it Evidence of a narrow structure in $\Upsilon (1S) l^+l^-$mass spectrum and CMS Phase I and II silicon detector upgrade studies},
\href{http://dx.doi.org/10.17077/etd.fxooie6m}{DOI: 10.17077/etd.fxooie6m}.
%0 citations counted in INSPIRE as of 13 May 2022

%\cite{ANDY:2019bfn}
\bibitem{ANDY:2019bfn}
L.~C.~Bland \textit{et al.} [ANDY Collaboration],
{\it Observation of Feynman scaling violations and evidence for a new resonance at RHIC},
\href{http://arxiv.org/abs/arXiv:1909.03124}{arXiv:1909.03124 [nucl-ex]}.
%14 citations counted in INSPIRE as of 17 May 2022

%\cite{Anwar:2017toa}
\bibitem{Anwar:2017toa}
M.~N.~Anwar, J.~Ferretti, F.~K.~Guo, E.~Santopinto and B.~S.~Zou,
{\it Spectroscopy and decays of the fully-heavy tetraquarks},
\href{http://dx.doi.org/10.1140/epjc/s10052-018-6073-9}{Eur. Phys. J. C \textbf{78} (2018) no.8, 647}.
%[arXiv:1710.02540 [hep-ph]].
%98 citations counted in INSPIRE as of 12 May 2022

%\cite{Esposito:2018cwh}
\bibitem{Esposito:2018cwh}
A.~Esposito and A.~D.~Polosa,
{\it A $bb\bar b\bar b$ di-bottomonium at the LHC?}
\href{http://dx.doi.org/10.1140/epjc/s10052-018-6269-z}{Eur. Phys. J. C \textbf{78} (2018) no.9, 782}.
%[arXiv:1807.06040 [hep-ph]].
%71 citations counted in INSPIRE as of 12 May 2022

%\cite{Hughes:2017xie}
\bibitem{Hughes:2017xie}
C.~Hughes, E.~Eichten and C.~T.~H.~Davies,
{\it Searching for beauty-fully bound tetraquarks using lattice nonrelativistic QCD},
\href{http://dx.doi.org/10.1103/PhysRevD.97.054505}{Phys. Rev. D \textbf{97} (2018) no.5, 054505}.
%[arXiv:1710.03236 [hep-lat]].
%75 citations counted in INSPIRE as of 12 May 2022

%\cite{Karliner:2016zzc}
\bibitem{Karliner:2016zzc}
M.~Karliner, S.~Nussinov and J.~L.~Rosner,
{\it $Q Q \bar Q \bar Q$ states: Masses, production, and decays},
\href{http://dx.doi.org/10.1103/PhysRevD.95.034011}{Phys. Rev. D \textbf{95} (2017) no.3, 034011}.
%[arXiv:1611.00348 [hep-ph]].
%125 citations counted in INSPIRE as of 12 May 2022

%\cite{Wu:2016vtq}
\bibitem{Wu:2016vtq}
J.~Wu, Y.~R.~Liu, K.~Chen, X.~Liu and S.~L.~Zhu,
{\it Heavy-flavored tetraquark states with the $QQ\bar{Q}\bar{Q}$ configuration},
\href{http://dx.doi.org/10.1103/PhysRevD.97.094015}{Phys. Rev. D \textbf{97} (2018) no.9, 094015}.
%[arXiv:1605.01134 [hep-ph]].
%109 citations counted in INSPIRE as of 12 May 2022

%\cite{Richard:2017vry}
\bibitem{Richard:2017vry}
J.~M.~Richard, A.~Valcarce and J.~Vijande,
{\it String dynamics and metastability of all-heavy tetraquarks},
\href{http://dx.doi.org/10.1103/PhysRevD.95.054019}{Phys. Rev. D \textbf{95} (2017) no.5, 054019}.
%[arXiv:1703.00783 [hep-ph]].
%79 citations counted in INSPIRE as of 12 May 2022

%\cite{Bai:2016int}
\bibitem{Bai:2016int}
Y.~Bai, S.~Lu and J.~Osborne,
{\it Beauty-full tetraquarks},
\href{http://dx.doi.org/10.1016/j.physletb.2019.134930}{Phys. Lett. B \textbf{798} (2019), 134930}.
%[arXiv:1612.00012 [hep-ph]].
%84 citations counted in INSPIRE as of 12 May 2022

%\cite{Chen:2019dvd}
\bibitem{Chen:2019dvd}
X.~Chen,
{\it Analysis of hidden-bottom $bb \bar{b} \bar{b}$ states},
\href{http://dx.doi.org/10.1140/epja/i2019-12807-2}{Eur. Phys. J. A \textbf{55} (2019) no.7, 106}.
%[arXiv:1902.00008 [hep-ph]].
%26 citations counted in INSPIRE as of 12 May 2022

%\cite{Debastiani:2017msn}
\bibitem{Debastiani:2017msn}
V.~R.~Debastiani and F.~S.~Navarra,
{\it A non-relativistic model for the $[cc][\bar{c}\bar{c}]$ tetraquark},
\href{http://dx.doi.org/10.1088/1674-1137/43/1/013105}{Chin. Phys. C \textbf{43} (2019) no.1, 013105}.
%[arXiv:1706.07553 [hep-ph]].
%89 citations counted in INSPIRE as of 12 May 2022

%\cite{Wang:2019rdo}
\bibitem{Wang:2019rdo}
G.~J.~Wang, L.~Meng and S.~L.~Zhu,
{\it Spectrum of the fully-heavy tetraquark state $QQ\bar Q' \bar Q'$},
\href{http://dx.doi.org/10.1103/PhysRevD.100.096013}{Phys. Rev. D \textbf{100} (2019) no.9, 096013}.
%[arXiv:1907.05177 [hep-ph]].
%59 citations counted in INSPIRE as of 12 May 2022

%\cite{Liu:2019zuc}
\bibitem{Liu:2019zuc}
M.~S.~Liu, Q.~F.~L\"u, X.~H.~Zhong and Q.~Zhao,
{\it All-heavy tetraquarks},
\href{http://dx.doi.org/10.1103/PhysRevD.100.016006}{Phys. Rev. D \textbf{100} (2019) no.1, 016006}.
%[arXiv:1901.02564 [hep-ph]].
%76 citations counted in INSPIRE as of 12 May 2022

%\cite{Deng:2020iqw}
\bibitem{Deng:2020iqw}
C.~Deng, H.~Chen and J.~Ping,
{\it Towards the understanding of fully-heavy tetraquark states from various models},
\href{http://dx.doi.org/10.1103/PhysRevD.103.014001}{Phys. Rev. D \textbf{103} (2021) no.1, 014001}.
%[arXiv:2003.05154 [hep-ph]].
%45 citations counted in INSPIRE as of 12 May 2022

%\cite{Richard:2018yrm}
\bibitem{Richard:2018yrm}
  J.~M.~Richard, A.~Valcarce and J.~Vijande,
  {\it Few-body quark dynamics for doubly heavy baryons and tetraquarks},
\href{http://dx.doi.org/10.1103/PhysRevC.97.035211}{Phys.\ Rev.\ C {\bf 97}, 035211 (2018)}.
  %%CITATION = doi:10.1103/PhysRevC.97.035211;%%
  %27 citations counted in INSPIRE as of 26 Jun 2020

%\cite{Wang:2017jtz}
\bibitem{Wang:2017jtz}
  Z.~G.~Wang,
  {\it Analysis of the $QQ\bar{Q}\bar{Q}$ tetraquark states with QCD sum rules},
\href{http://dx.doi.org/10.1140/epjc/s10052-017-4997-0}{Eur.\ Phys.\ J.\ C {\bf 77}, 432 (2017)}.
  %%CITATION = doi:10.1140/epjc/s10052-017-4997-0;%%
  %44 citations counted in INSPIRE as of 26 Jun 2020

%\cite{Wang:2018poa}
\bibitem{Wang:2018poa}
  Z.~G.~Wang and Z.~Y.~Di,
  {\it Analysis of the Vector and Axialvector $QQ\bar{Q}\bar{Q}$ Tetraquark States with QCD Sum Rules},
\href{http://dx.doi.org/10.5506/APhysPolB.50.1335}{Acta Phys.\ Polon.\ B {\bf 50}, 1335 (2019)}.
  %%CITATION = doi:10.5506/APhysPolB.50.1335;%%
  %11 citations counted in INSPIRE as of 27 Jul 2020

%\cite{Chiu:2007km}
\bibitem{Chiu:2007km}
T.~W.~Chiu \textit{et al.} [TWQCD Collaboration],
{\it Beauty mesons in lattice QCD with exact chiral symmetry},
\href{http://dx.doi.org/10.1016/j.physletb.2007.06.017}{Phys. Lett. B \textbf{651} (2007), 171-176}.
%[arXiv:0705.2797 [hep-lat]].
%83 citations counted in INSPIRE as of 09 May 2022

%\cite{LHCb:2018uwm}
\bibitem{LHCb:2018uwm}
R.~Aaij \textit{et al.} [LHCb Collaboration],
{\it Search for beautiful tetraquarks in the $\Upsilon(1S)\mu^+\mu^-$ invariant-mass spectrum},
\href{http://dx.doi.org/10.1007/JHEP10(2018)086}{JHEP \textbf{10} (2018), 086}.
%[arXiv:1806.09707 [hep-ex]].
%55 citations counted in INSPIRE as of 13 May 2022

%\cite{CMS:2020qwa}
\bibitem{CMS:2020qwa}
A.~M.~Sirunyan \textit{et al.} [CMS Collaboration],
{\it Measurement of the $\Upsilon$(1S) pair production cross section and search for resonances decaying to $\Upsilon$(1S)$\mu^+\mu^-$ in proton-proton collisions at $\sqrt{s} =$ 13 TeV},
\href{http://dx.doi.org/10.1016/j.physletb.2020.135578}{Phys. Lett. B \textbf{808} (2020), 135578}.
%[arXiv:2002.06393 [hep-ex]].
%40 citations counted in INSPIRE as of 13 May 2022

%\cite{Chen:2020xwe}
\bibitem{Chen:2020xwe}
H.~X.~Chen, W.~Chen, X.~Liu and S.~L.~Zhu,
{\it Strong decays of fully-charm tetraquarks into di-charmonia},
\href{http://dx.doi.org/10.1016/j.scib.2020.08.038}{Sci. Bull. \textbf{65} (2020), 1994-2000}.
%[arXiv:2006.16027 [hep-ph]].
%51 citations counted in INSPIRE as of 26 Apr 2022

%\cite{Chen:2016jxd}
\bibitem{Chen:2016jxd}
  W.~Chen, H.~X.~Chen, X.~Liu, T.~G.~Steele and S.~L.~Zhu,
  {\it Hunting for exotic doubly hidden-charm/bottom tetraquark states},
\href{http://dx.doi.org/10.1016/j.physletb.2017.08.034}{Phys.\ Lett.\ B {\bf 773}, 247 (2017)}.
  %%CITATION = doi:10.1016/j.physletb.2017.08.034;%%
  %39 citations counted in INSPIRE as of 22 Jun 2020

%\cite{Chen:2019wjd}
\bibitem{Chen:2019wjd}
H.~X.~Chen,
{\it Decay properties of the $Z_c(3900)$ through the Fierz rearrangement},
\href{http://dx.doi.org/10.1088/1674-1137/abae4b}{Chin. Phys. C \textbf{44} (2020) no.11, 114003}.
%[arXiv:1910.03269 [hep-ph]].
%7 citations counted in INSPIRE as of 14 May 2022

%\cite{Chen:2019eeq}
\bibitem{Chen:2019eeq}
H.~X.~Chen,
{\it Decay properties of the X(3872) through the Fierz rearrangement},
\href{http://dx.doi.org/10.1088/1572-9494/ac449e}{Commun. Theor. Phys. \textbf{74} (2022) no.2, 025201}.
%[arXiv:1911.00510 [hep-ph]].
%4 citations counted in INSPIRE as of 14 May 2022

%\cite{Chen:2020pac}
\bibitem{Chen:2020pac}
H.~X.~Chen,
{\it Decay properties of $P_c$ states through the Fierz rearrangement},
\href{http://dx.doi.org/10.1140/epjc/s10052-020-08519-1}{Eur. Phys. J. C \textbf{80} (2020) no.10, 945}.
%[arXiv:2001.09563 [hep-ph]].
%10 citations counted in INSPIRE as of 14 May 2022

%\cite{Chen:2020opr}
\bibitem{Chen:2020opr}
H.~X.~Chen,
{\it Hidden-charm pentaquark states through current algebra: from their production to decay},
\href{http://dx.doi.org/10.1088/1674-1137/ac6ed2}{Chin. Phys. C \textbf{46} (2022) no.9, 093105}.

%\cite{Chen:2021erj}
\bibitem{Chen:2021erj}
H.~X.~Chen,
{\it Hadronic molecules in B decays},
\href{http://dx.doi.org/10.1103/PhysRevD.105.094003}{Phys. Rev. D \textbf{105} (2022) no.9, 094003}.
%[arXiv:2103.08586 [hep-ph]].
%20 citations counted in INSPIRE as of 14 May 2022

%\cite{Dias:2013xfa}
\bibitem{Dias:2013xfa}
J.~M.~Dias, F.~S.~Navarra, M.~Nielsen and C.~M.~Zanetti,
{\it $Z^+_c$(3900) decay width in QCD sum rules},
\href{http://dx.doi.org/10.1103/PhysRevD.88.016004}{Phys. Rev. D \textbf{88} (2013) no.1, 016004}.
%101 citations counted in INSPIRE as of 18 Oct 2022

%\cite{Agaev:2016dev}
\bibitem{Agaev:2016dev}
S.~S.~Agaev, K.~Azizi and H.~Sundu,
{\it Strong $Z_c^{+}(3900)\rightarrow J/\psi \pi^{+}; \eta_{c} \rho^{+}$ decays in QCD},
\href{http://dx.doi.org/10.1103/PhysRevD.93.074002}{Phys. Rev. D \textbf{93} (2016) no.7, 074002}.

%\cite{Esposito:2014hsa}
\bibitem{Esposito:2014hsa}
A.~Esposito, A.~L.~Guerrieri and A.~Pilloni,
{\it Probing the nature of $Z_c^{(\prime)}$ states via the $\eta_c \rho$ decay},
\href{http://dx.doi.org/10.1016/j.physletb.2015.04.057}{Phys. Lett. B \textbf{746} (2015), 194-201}.

%\cite{Voloshin:2019aut}
\bibitem{Voloshin:2019aut}
  M.~B.~Voloshin,
  {\it Some decay properties of hidden-charm pentaquarks as baryon-meson molecules},
  \href{http://dx.doi.org/10.1103/PhysRevD.100.034020}{Phys.\ Rev.\ D {\bf 100}, 034020 (2019)}.
  %%CITATION = doi:;%%
  %15 citations counted in INSPIRE as of 21 Jan 2020

%\cite{Voloshin:2013dpa}
\bibitem{Voloshin:2013dpa}
  M.~B.~Voloshin,
  {\it $Z_c(3900)$ - what is inside?}
  \href{http://dx.doi.org/10.1103/PhysRevD.87.091501}{Phys.\ Rev.\ D {\bf 87}, 091501(R) (2013)}.
  %%CITATION = doi:;%%
  %91 citations counted in INSPIRE as of 22 Sep 2019

%\cite{Maiani:2017kyi}
\bibitem{Maiani:2017kyi}
  L.~Maiani, A.~D.~Polosa and V.~Riquer,
  {\it A theory of $X$ and $Z$ multiquark resonances},
  \href{http://dx.doi.org/10.1016/j.physletb.2018.01.039}{Phys.\ Lett.\ B {\bf 778}, 247 (2018)}.
  %%CITATION = doi:;%%
  %24 citations counted in INSPIRE as of 30 Sep 2019

%\cite{Wang:2019spc}
\bibitem{Wang:2019spc}
G.~J.~Wang, L.~Y.~Xiao, R.~Chen, X.~H.~Liu, X.~Liu and S.~L.~Zhu,
{\it Probing hidden-charm decay properties of $P_c$ states in a molecular scenario},
\href{http://dx.doi.org/10.1103/PhysRevD.102.036012}{Phys. Rev. D \textbf{102} (2020) no.3, 036012}.

%\cite{Xiao:2019spy}
\bibitem{Xiao:2019spy}
L.~Y.~Xiao, G.~J.~Wang and S.~L.~Zhu,
{\it Hidden-charm strong decays of the $Z_c$ states},
\href{http://dx.doi.org/10.1103/PhysRevD.101.054001}{Phys. Rev. D \textbf{101} (2020) no.5, 054001}.

%\cite{Cheng:2020nho}
\bibitem{Cheng:2020nho}
J.~B.~Cheng, S.~Y.~Li, Y.~R.~Liu, Y.~N.~Liu, Z.~G.~Si and T.~Yao,
{\it Spectrum and rearrangement decays of tetraquark states with four different flavors},
\href{http://dx.doi.org/10.1103/PhysRevD.101.114017}{Phys. Rev. D \textbf{101} (2020) no.11, 114017}.

%\cite{Su:2022eun}
\bibitem{Su:2022eun}
N.~Su and H.~X.~Chen,
{\it S- and P-wave fully strange tetraquark states from QCD sum rules},
\href{http://dx.doi.org/10.1103/PhysRevD.106.014023}{Phys. Rev. D \textbf{106} (2022) no.1, 014023}.

%\cite{Chen:2020aos}
\bibitem{Chen:2020aos}
H.~X.~Chen, W.~Chen, R.~R.~Dong and N.~Su,
{\it $X_0$(2900) and $X_1$(2900): Hadronic Molecules or Compact Tetraquarks},
\href{http://dx.doi.org/10.1088/0256-307X/37/10/101201}{Chin. Phys. Lett. \textbf{37} (2020) no.10, 101201}.

%\cite{Chen:2020uif}
\bibitem{Chen:2020uif}
H.~X.~Chen, W.~Chen, X.~Liu and X.~H.~Liu,
{\it Establishing the first hidden-charm pentaquark with strangeness},
\href{http://dx.doi.org/10.1140/epjc/s10052-021-09196-4}{Eur. Phys. J. C \textbf{81} (2021) no.5, 409}.

%\cite{Liu:2020lpw}
\bibitem{Liu:2020lpw}
F.~X.~Liu, M.~S.~Liu, X.~H.~Zhong and Q.~Zhao,
{\it Fully strange tetraquark $ss\bar{s}\bar{s}$ spectrum and possible experimental evidence},
\href{http://dx.doi.org/10.1103/PhysRevD.103.016016}{Phys. Rev. D \textbf{103}, 016016 (2021)}.
%[arXiv:2008.01372 [hep-ph]].
%11 citations counted in INSPIRE as of 20 Apr 2022

%\cite{Veliev:2010gb}
\bibitem{Veliev:2010gb}
E.~V.~Veliev, H.~Sundu, K.~Azizi and M.~Bayar,
{\it Scalar quarkonia at finite temperature},
\href{http://dx.doi.org/10.1103/PhysRevD.82.056012}{Phys. Rev. D \textbf{82} (2010), 056012}.
%[arXiv:1003.0119 [hep-ph]].
%40 citations counted in INSPIRE as of 09 May 2022

%\cite{Becirevic:2013bsa}
\bibitem{Becirevic:2013bsa}
D.~Be\v{c}irevi\'c, G.~Duplan\v{c}i\'c, B.~Klajn, B.~Meli\'c and F.~Sanfilippo,
{\it Lattice QCD and QCD sum rule determination of the decay constants of $\eta_c$, J/$\psi$ and $h_c$ states},
\href{http://dx.doi.org/10.1016/j.nuclphysb.2014.03.024}{Nucl. Phys. B \textbf{883} (2014), 306-327}.
%[arXiv:1312.2858 [hep-ph]].
%85 citations counted in INSPIRE as of 13 May 2022

%\cite{MaiordeSousa:2012vv}
\bibitem{MaiordeSousa:2012vv}
M.~S.~Maior de Sousa and R.~R.~da Silva,
{\it The $\rho(2S)$, $\psi(2S)$ and $\Upsilon(2S)$ and $\psi_t(1S,2S)$ Mesons in a Double Pole QCD Sum Rules},
\href{http://dx.doi.org/10.1007/s13538-016-0449-9}{Braz. J. Phys. \textbf{46} (2016) no.6, 730-739}.
%[arXiv:1205.6793 [hep-ph]].
%18 citations counted in INSPIRE as of 09 May 2022

%\cite{Novikov:1977dq}
\bibitem{Novikov:1977dq}
V.~A.~Novikov, L.~B.~Okun, M.~A.~Shifman, A.~I.~Vainshtein, M.~B.~Voloshin and V.~I.~Zakharov,
{\it Charmonium and gluons},
\href{http://dx.doi.org/10.1016/0370-1573(78)90120-5}{Phys. Rept. \textbf{41} (1978), 1-133}.
%903 citations counted in INSPIRE as of 13 May 2022

%\cite{Beneke:1999br}
\bibitem{Beneke:1999br}
  M.~Beneke, G.~Buchalla, M.~Neubert and C.~T.~Sachrajda,
  {\it QCD Factorization for $B \to \pi \pi$ Decays: Strong Phases and CP Violation in the Heavy Quark Limit},
\href{http://dx.doi.org/10.1103/PhysRevLett.83.1914}{Phys.\ Rev.\ Lett.\  {\bf 83}, 1914 (1999)}.
  %%CITATION = doi:10.1103/PhysRevLett.83.1914;%%
  %1252 citations counted in INSPIRE as of 12 Sep 2019

%\cite{Beneke:2000ry}
\bibitem{Beneke:2000ry}
  M.~Beneke, G.~Buchalla, M.~Neubert and C.~T.~Sachrajda,
  {\it QCD factorization for exclusive, nonleptonic $B$-meson decays: general arguments and the case of heavy-light final states},
\href{http://dx.doi.org/10.1016/S0550-3213(00)00559-9}{Nucl.\ Phys.\ B {\bf 591}, 313 (2000)}.
  %%CITATION = doi:10.1016/S0550-3213(00)00559-9;%%
  %1190 citations counted in INSPIRE as of 12 Sep 2019

%\cite{Beneke:2001ev}
\bibitem{Beneke:2001ev}
  M.~Beneke, G.~Buchalla, M.~Neubert and C.~T.~Sachrajda,
  {\it QCD factorization in $B \to \pi K$, $\pi \pi$ decays and extraction of Wolfenstein parameters},
\href{http://dx.doi.org/10.1016/S0550-3213(01)00251-6}{Nucl.\ Phys.\ B {\bf 606}, 245 (2001)}.
  %%CITATION = doi:10.1016/S0550-3213(01)00251-6;%%
  %1022 citations counted in INSPIRE as of 21 Sep 2019

%\cite{Li:2020rcg}
\bibitem{Li:2020rcg}
  H.~D.~Li, C.~D.~L{\"u}, C.~Wang, Y.~M.~Wang and Y.~B.~Wei,
  {\it QCD calculations of radiative heavy meson decays with subleading power corrections},
\href{http://dx.doi.org/10.1007/JHEP04(2020)023}{JHEP {\bf 2004}, 023 (2020)}.
  %%CITATION = doi:10.1007/JHEP04(2020)023;%%
  %1 citations counted in INSPIRE as of 12 Jun 2020

%\cite{Yang:2020zrh}
\bibitem{Yang:2020zrh}
H.~M.~Yang and H.~X.~Chen,
{\it $P$-wave bottom baryons of the $SU(3)$ flavor $\mathbf{6}_F$},
\href{http://dx.doi.org/10.1103/PhysRevD.101.114013}{Phys. Rev. D \textbf{101} (2020) no.11, 114013}
[erratum: Phys. Rev. D \textbf{102} (2020) no.7, 079901].

%\cite{Yang:2021lce}
\bibitem{Yang:2021lce}
H.~M.~Yang and H.~X.~Chen,
{\it $P$-wave charmed baryons of the $SU(3)$ flavor $6_F$},
\href{http://dx.doi.org/10.1103/PhysRevD.104.034037}{Phys. Rev. D \textbf{104} (2021) no.3, 034037}.


\end{thebibliography}
\end{document}